\documentclass[a4paper,leqno]{article}

\usepackage[english]{babel}
\usepackage{graphicx}
\usepackage{hyperref}
\usepackage{bbold}
\usepackage{url}
\usepackage{amsmath}
\usepackage{amsfonts}
\usepackage{color}
\usepackage{verbatim}
\usepackage[utf8]{inputenc}
\usepackage{listings}

\makeatletter
\newcommand{\toto@c@page}{}
\newcommand{\toto@addmarginpar}{}
\let\toto@addmarginpar\@addmarginpar
\renewcommand{\@addmarginpar}{
  \def\toto@c@page{\c@page}
  \c@page=1
  \toto@addmarginpar
  \c@page=\toto@c@page
}
\makeatother
\usepackage[textwidth=5cm,textsize=footnotesize]{todonotes}

\newcommand{\code}{\sf}
\newcommand{\codelisting}{\code}

\lstdefinelanguage{Coq}{
  comment=[n]{(*}{*)},
  otherkeywords={->, /\\, \\/, =>},
  morekeywords={Goal, Proof, Definition, Theorem, Lemma, Qed, Time, Require, Notation, Variable, Section, End, Fixpoint, forall, exists, fun, let, match, with, end},
  literate={=>}{{$\Rightarrow$}}1 {->}{{$\to$}}1 {/\\}{{$\land$}}1 {<=}{{$\le$}}1 {>=}{{$ge$}}1
}

\newcommand{\derparun}[2]{\frac{\partial {#1}}{\partial {#2}}}
\newcommand{\derpar}[3]{\frac{\partial^{#1} {#2}}{\partial {#3}^{#1}}}
\newcommand{\eqdef}{{\;\stackrel{\text{def}}{=}\;}}
\newcommand{\floor}[1]{\left\lfloor#1\right\rfloor}

\newcommand{\Z}{\mathbb{Z}}

\newcommand{\R}{{\mathbb R}}

\newcommand{\demi}{\frac{1}{2}}
\newcommand{\prodscal}[3]{\left<#2,#3\right>_{#1}}
\newcommand{\norme}[2]{\left\|#2\right\|_{#1}}
\newcommand{\ps}[2]{\prodscal{}{#1}{#2}}
\newcommand{\n}[1]{\norme{}{#1}}
\newcommand{\nA}[1]{\norme{A(c)}{#1}}
\newcommand{\psdx}[2]{\prodscal{\Delta x}{#1}{#2}}
\newcommand{\ndx}[1]{\norme{\Delta x}{#1}}
\newcommand{\psAh}[2]{\prodscal{A_h(c)}{#1}{#2}}
\newcommand{\nAh}[1]{\norme{A_h(c)}{#1}}

\newcommand{\norm}[1]{\ensuremath{\|#1\|}}

\newcommand{\xmin}{x_{\rm min}}
\newcommand{\xmax}{x_{\rm max}}
\newcommand{\tmax}{t_{\rm max}}
\newcommand{\imax}{i_{\rm max}}
\newcommand{\kmax}{k_{\rm max}}
\newcommand{\bfx}{{\bf x}}
\newcommand{\bfdeltax}{{\bf \Delta x}}
\newcommand{\eps}{\varepsilon}

\newcommand{\soft}[1]{{\sf #1}}

\newcommand{\lang}[1]{\soft{#1}}

\newcommand{\clang}{\lang{C}}

\newcommand{\Fost}{F$\!\oint$st}

\newcommand{\ANRCerPanetFost}{This research was supported by the ANR projects
  CerPAN (ANR-05-BLAN-0281-04) and
  {\Fost} (ANR-08-BLAN-0246-01).}

\def\adots{\mathinner{\mkern2mu\raise 1pt\hbox{.}\mkern 3mu\raise 4pt\hbox{.}\mkern1mu\raise 7pt\hbox{{.}}}}
\newcommand{\ra}{\rightarrow}

\usepackage{RRA4}
\usepackage{hyperref}
\usepackage{a4wide}

\RRdate{D\'ecembre 2011}

\RRauthor{
  Sylvie Boldo%
  \thanks[1]{Projet ProVal.
    {\tt \{Sylvie.Boldo,Jean-Christophe.Filliatre,Guillaume.Melquiond\}%
      @inria.fr}.}%
  \thanks[2]{LRI, UMR 8623, Universit\'e Paris-Sud, CNRS, Orsay cedex,
    F-91405.}%
  \and Fran\c{c}ois Cl\'ement%
  \thanks[3]{Projet Pomdapi. {\tt \{Francois.Clement,Pierre.Weis\}@inria.fr}.}%
  \and Jean-Christophe Filli\^atre%
  \thanksref{2}%
  \thanksref{1}%
  \and Micaela Mayero%
  \thanks[4]{LIPN, UMR 7030, Universit\'e Paris-Nord, CNRS, Villetaneuse,
    F-93430.\goodbreak
    {\tt Micaela.Mayero@lipn.univ-paris13.fr}.}%
  \thanks[5]{LIP, Ar\'enaire (INRIA Grenoble - Rh\^one-Alpes, CNRS UMR 5668,
  UCBL, ENS Lyon), Lyon, F-69364.}%
  \and Guillaume Melquiond%
  \thanksref{1}%
  \thanksref{2}%
  \and Pierre Weis%
  \thanksref{3}
}
\authorhead{S. Boldo, F. Cl\'ement, J-C. Filli\^atre, M. Mayero, G. Melquiond,
  \& P. Weis}

\RRtitle{R\'esolution num\'erique de l'\'equation des ondes : une preuve
  m\'ecanis\'ee compl\`ete d'un programme C}
\RRetitle{Wave Equation Numerical Resolution: a Comprehensive Mechanized
  Proof of a C Program}

\RRnote{\ANRCerPanetFost}

\RRresume{Nous prouvons formellement la correction d'un programme C impl\'ementant un
sch\'ema num\'erique pour la r\'esolution de l'\'equation des ondes
acoustiques en dimension~1.
Une telle impl\'ementation introduit diff\'erents types d'erreurs~: l'erreur de
m\'ethode due au sch\'ema num\'erique et l'erreur d'arrondi due aux calculs
en virgule flottante.
Nous annotons ce programme C pour sp\'ecifier ces deux types d'erreur.
Nous utilisons Frama-C pour g\'en\'erer les th\'eor\`emes qui garantissent la
correction du code.
Nous prouvons ces th\'eor\`emes \`a l'aide de solveurs SMT, de Gappa et de
Coq.
Un d\'eveloppement Coq important est n\'ecessaire pour prouver l'ad\'equation
du programme C au sch\'ema num\'erique et pour borner les erreurs.
\`A notre connaissance, c'est la premi\`ere fois qu'un tel programme d'analyse
num\'erique est compl\`etement v\'erifi\'e m\'ecaniquement.
}
\RRabstract{We formally prove correct a C program that implements a
numerical scheme for the resolution of the one-dimensional acoustic wave
equation. Such an implementation introduces errors at several levels:
the numerical scheme introduces method errors, and floating-point
computations lead to round-off errors.  We annotate this C program to
specify both method error and round-off error.  We use Frama-C to
generate theorems that guarantee the soundness of the code. We
discharge these theorems using SMT solvers, Gappa, and Coq.  This
involves a large Coq development to prove the adequacy of the C
program to the numerical scheme and to bound errors.  To our
knowledge, this is the first time such a numerical analysis program
is fully machine-checked.
}

\RRmotcle{preuve formelle d'un programme num\'erique,
convergence d'un sch\'ema num\'erique,
preuve de programme C,
preuve formelle en Coq,
\'equation des ondes acoustiques,
\'equation aux d\'eriv\'ees partielles,
analyse d'erreur d'arrondi.
}
\RRkeyword{Formal proof of numerical program
\and Convergence of numerical scheme
\and Proof of C program
\and Coq formal proof
\and Acoustic wave equation
\and Partial differential equation
\and Rounding error analysis
}

\RRprojets{ProVal et Estime}

\RRdomaine{2 et 5}
\RRtheme{Programmation, v\'erification et preuves\\
  Observation et mod\'elisation pour les sciences de l'environnement}

\RCSaclay

\usepackage{amsthm}
\newtheorem{lemma}{Lemma}
\newtheorem{theorem}{Theorem}

\newenvironment{myproof}{\begin{proof}}{\end{proof}}
\newcommand{\mymultlinenewline}{\null}

\begin{document}

\RRNo{7826}
\makeRR

\section{Introduction}
\label{sec:introduction}

Ordinary differential equations (ODE) and partial differential
equations (PDE) are ubiquitous in engineering and scientific
computing. They show up in nuclear simulation, weather forecast, and
more generally in numerical simulation, including block diagram
modelization. Since solutions to nontrivial problems are non-analytic,
they must be approximated by numerical schemes over discrete grids.

Numerical analysis is a part of applied mathematics that is mainly interested
in proving the {\em convergence} of these schemes~\cite{cfl:pde:67}, that is,
proving that approximation quality increases as the size of discretization
steps decreases. The approximation quality represents the distance between
the exact continuous solution and the approximated discrete solution; this
distance must tend toward zero in order for the numerical scheme to be
useful.

A numerical scheme is typically proved to be convergent with pen and
paper. This is a difficult, time-consuming, and error-prone task.  Then the
scheme is implemented as a C/C++ or Fortran program. This introduces new
issues.  First, we must ensure that the program correctly implements the
scheme and is immune from runtime errors such as out-of-bounds accesses or
overflows. Second, the program introduces round-off errors due to
floating-point computations and we must prove that those errors could not
lead to irrelevant results. Typical pen-and-paper proofs do not address
floating-point nor runtime errors. Indeed the huge number of proof
obligations, and their complexity, make the whole process almost intractable.
However, with the help of mechanized program verification, such a proof
becomes feasible. In the first place, because automated theorem provers can
alleviate the proof burden. More importantly, because the proof is guaranteed
to cover all aspects of the verification.

\paragraph{Our case study.}

We consider the acoustic wave equation in an one-dimensional space domain.
The equation describes the propagation of pressure variations (or sound
waves) in a fluid medium; it also models the behavior of a vibrating string.
Among the wide variety of numerical schemes to approximate the 1D acoustic
wave equation, we choose the simplest one: the second order centered finite
difference scheme, also known as \emph{three-point scheme}.  To keep it
simple, we assume an homogeneous medium (the propagation velocity is
constant) and we consider discretization over regular grids with constant
discretization steps for time and space. Our goal is to prove the correctness
of a C program implementing this scheme.

\paragraph{Method and tools.}

We use the Jessie plug-in of Frama-C~\cite{marche07plpv,Frama-C} to
perform the deductive verification of this C program. The source code
is augmented with \emph{ACSL annotations}~\cite{ACSL} to describe its formal
specification. When submitted to Frama-C, proof obligations are
generated. Once these theorems are proved, the program is guaranteed
to satisfy its specification and to be free from runtime errors.
Part of the proof obligations are discharged by automated provers,
\emph{e.g.} Alt-Ergo~\cite{alt-ergo}, CVC3~\cite{CVC3},
Gappa~\cite{DauMel10}, and Z3~\cite{Z3}. The more complicated ones,
such as the one related to the convergence of the numerical scheme,
cannot be proved automatically.
These obligations were manually proved with the Coq~\cite{Coq,Coq-ref}
interactive proof assistant. In the end, we have formally verified all
the properties of the C program. To our knowledge, this is the first
time this kind of verification is machine-checked. The annotated C
program and the Coq sources of the formal development are available
from

\centerline{\url{http://fost.saclay.inria.fr/wave_total_error.html}}

\paragraph{State of the art.}

There is an abundant literature about the convergence of numerical
schemes, {\em e.g.} see~\cite{tho:npd:95,zwi:hde:98}. In particular, the
convergence of the three-point scheme for the wave equation is well-known
and taught relatively early~\cite{bec:esn:09}. Unfortunately, no article
goes into all the details needed for a formal proof. These mathematical
``details'' may have been skipped for readability, but some mandatory
details may have also been omitted due to oversights.

In the fields of automatic provers and proof assistants, few works
have been done for the formalization and mechanical proofs of
mathematical analysis, and even fewer works for numerical analysis.
The first developments on real numbers and real analysis are from the
late 90's, in systems such as ACL2~\cite{GK01}, Coq~\cite{May01}, HOL
Light~\cite{Har98}, Isabelle~\cite{Fle00}, Mizar~\cite{Rud92}, and
PVS~\cite{Dut96}.  An extensive work has been done by Harrison
regarding $\R^n$ and the dot product~\cite{Har05}. Constructive real
analysis~\cite{NiqGeu00,Cru02,KS11} and further developments in numerical
analysis~\cite{OC08,OCS10} have been carried out at Nijmegen.
We can also mention the formal proof of an automatic differentiation
algorithm~\cite{May02}.

As explained by Rosinger in 1985, old methods to bound round-off errors were
based on ``unrealistic linearizing assumptions''~\cite{Ros85}. Further work
was done under more realistic assumptions about round-off
errors~\cite{Ros85,Ros91}, but none of these assumptions were proved correct
with respect to the numerical schemes.  As Roy and Oberkampf, we believe that
round-off errors should not be treated as random variables and that
traditional statistical methods should not be used~\cite{RoOb11}. They
propose the use of interval arithmetic or increased precision to control
accuracy. Note that their example of hypersonic nozzle flow is converging so
fast that round-off errors can be neglected~\cite{RoOb11}.  Interval
arithmetic can also take method error into account~\cite{Szy09}. The final
interval is then claimed to contain the exact solution. This is not formally
proved, though.  Additionally, the width of the final interval can be quite
large.

There are other tools to bound round-off errors not dedicated to
numerical schemes.  Some successful approaches are based on abstract
interpretation~\cite{ASTREE_ESOP05,DGPSTV09}.  In our case, they are
of little help, since there is a complex phenomenon of error
compensation during the computations. Ignoring this compensation would
lead to bounds on round-off errors growing as fast as $O(2^k)$ ($k$
being the number of time steps). That is why we had to thoroughly
study the propagation of round-off errors in this numerical scheme
to get tighter bounds. It also means that most of the proofs had
to be done by hand to achieve this part of the formal verification.

\paragraph{Outline.}

Section~\ref{sec:wave} presents the PDE, the numerical scheme, and their
mathematical properties.
Section~\ref{sec:errors} is devoted to the proofs of the
convergence of the numerical scheme and the upper bound for the round-off error.
Finally, Section~\ref{sec:mechanization} describes the
formalization, \emph{i.e.} the tools used, the annotated C program, and
the mechanized proofs.



\section[Numerical scheme for the wave equation]{Numerical Scheme for the Wave Equation}
\label{sec:wave}

A partial differential equation (PDE) modeling an evolution problem is an
equation involving partial derivatives of an unknown function of several
independent space and time variables.
The uniqueness of the solution is obtained by imposing initial conditions,
i.e. values of the function and some of its derivatives at
initial time.
The problem of the vibrating string tied down at both ends, among many other
physical problems, is formulated as an {\em initial-boundary value problem}
where the boundary conditions are additional constraints set on the boundary
of the supposedly bounded domain~\cite{tho:npd:95}.

This section, as well as the steps taken at Section~\ref{sec:wave_proof} to
conduct the convergence proof of the numerical scheme, is
inspired by~\cite{bec:esn:09}.

\subsection[The continuous equation]{The Continuous Equation}
\label{sec:continuous}

\begin{figure}[ht]
  \begin{center}
    \includegraphics[width=12cm]{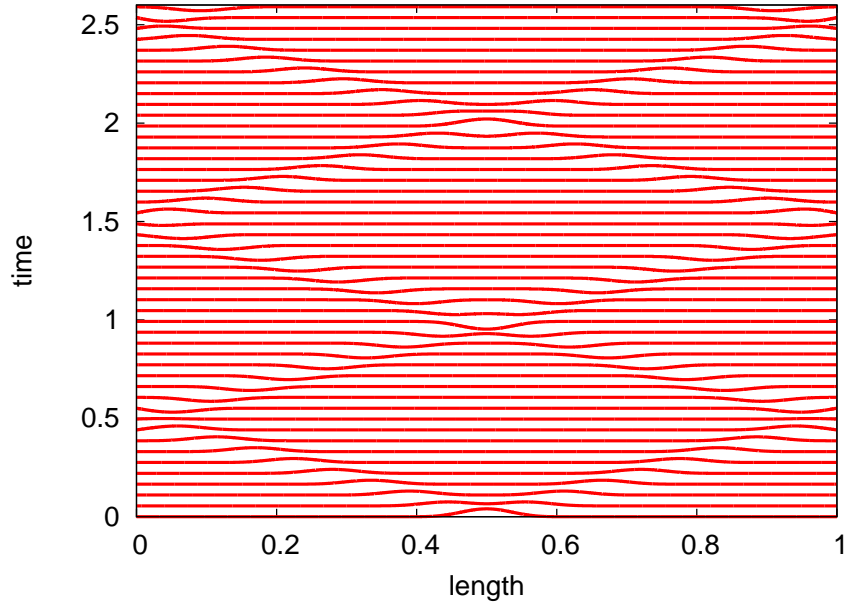}
  \end{center}
  \caption{Space-time representation of the signal propagating along a
    vibrating string.
    Each curve represents the string at a different time step.}
  \label{f:waves}
\end{figure}

The chosen PDE models the propagation of waves along an ideal vibrating elastic
string that is tied down at both ends, see~\cite{ach:wpe:73,bg:mcw:94}, see
also Figure~\ref{f:waves}.
The PDE is obtained from Newton's laws of motion~\cite{new:alm:87}.

The gravity is neglected, so the string is supposed rectilinear when at rest.
Let~$\xmin$ and~$\xmax$ be the abscissas of the extremities of the string.
Let $\Omega=[\xmin,\xmax]$ be the bounded space domain.
Let~$p(x,t)$ be the transverse displacement of the point of the string of
abscissa~$x$ at time~$t$ from its equilibrium position; it is a (signed)
scalar.
Let~$c$ be the constant propagation velocity; it is a positive number that
depends on the section and density of the string.
Let~$s(x,t)$ be the external action on the point of abscissa~$x$ at
time~$t$; it is a source term, such that $t=0\Rightarrow s(x,t)=0$.
Finally, let~$p_0(x)$ and~$p_1(x)$ be the initial position and velocity of
the point of abscissa~$x$.
We consider the initial-boundary value problem
\begin{eqnarray}
  \label{e:L}
  \forall t \ge 0,\; \forall x \in \Omega, & &
  (L (c) \, p) (x, t) \eqdef
  \derpar{2}{p}{t} (x, t) + A (c) \, p (x, t) =
  s (x, t), \\
  \label{e:L1}
  \forall x \in \Omega, & &
  (L_1 \, p) (x, 0) \eqdef
  \derparun{p}{t} (x, 0) =
  p_1 (x), \\
  \label{e:L0}
  \forall x \in \Omega, & &
  (L_0 \, p) (x, 0) \eqdef
  p (x, 0) =
  p_0 (x), \\
  \label{e:dir}
  \forall t \ge 0, & &
  p (\xmin, t) =
  p (\xmax, t) =
  0
\end{eqnarray}
where the differential operator~$A(c)$ is defined by
\begin{equation}
  \label{e:A}
  A (c) \eqdef - c^2 \derpar{2}{}{x}.
\end{equation}

This simple partial derivative equation happens to possess an analytical
solution given by the so-called d'Alembert's formula~\cite{dal:rcf:47},
obtained from the method of characteristics and the image
theory~\cite{joh:pde:86}, $\forall t\ge 0$, $\forall x\in\Omega$,
\begin{multline}
  \label{e:dAlembert}
  p (x, t) =
  \demi (\tilde{p}_0 (x - ct) + \tilde{p}_0 (x + ct)) +
  \frac{1}{2c} \int_{x - ct}^{x + ct} \tilde{p}_1 (y) dy + \mymultlinenewline
  \frac{1}{2c} \int_0^t \left(
    \int_{x - c(t - \sigma)}^{x + c(t - \sigma)} \tilde{s} (y, \sigma) dy
  \right) d\sigma
\end{multline}
where the quantities~$\tilde{p}_0$, $\tilde{p}_1$, and~$\tilde{s}$ are
respectively the successive antisymmetric extensions in space of~$p_0$, $p_1$,
and~$s$ defined on $\Omega$ to the entire real axis~$\R$.

We have formally verified d'Alembert's formula as a separate
work~\cite{LelMel12}.
But for the purpose of the current work, we just admit
that under reasonable conditions on the Cauchy data~$p_0$ and~$p_1$
and on the source term~$s$, there exists a unique solution~$p$ to the
initial-boundary value problem~(\ref{e:L})--(\ref{e:dir}) for each $c>0$.
Simply supposing the existence of a solution instead of
exhibiting it, opens the way to scale our approach to more
general cases for which there is no known analytic expression of a solution,
\emph{e.g.} in the case of a nonuniform propagation velocity~$c$.

For such a solution~$p$, it is natural to associate at each time~$t$ the
positive definite quadratic quantity
\begin{equation}
  \label{e:energiecontinue}
  E (c) (p) (t) \eqdef
  \demi \n{\left(x \mapsto \derparun{p}{t} (x, t)\right)}^2 +
  \demi \nA{(x \mapsto p (x, t))}^2
\end{equation}
where $\ps{q}{r}\eqdef\int_\Omega q(x)r(x)dx$,
$\n{q}^2\eqdef\ps{q}{q}$ and $\nA{q}^2\eqdef\ps{A(c)\,q}{q}$.
The first term is interpreted as the kinetic energy, and the second term as the
potential energy, making~$E$ the mechanical energy of the vibrating string.

\subsection[The discrete equations]{The Discrete Equations}
\label{sec:discrete}

Let~$\imax$ be the positive number of intervals of the space discretization.
Let the space discretization step $\Delta x$ and the discretization function
$i_{\Delta x}$ be defined as
\[\Delta x\eqdef\frac{\xmax-\xmin}{\imax} \quad \mbox{and} \quad
i_{\Delta x}(x)\eqdef\floor{\frac{x-\xmin}{\Delta x}}.\]

Let us consider the time interval $[0,\tmax]$.
Let $\Delta t\in]0,\tmax[$ be the time discretization step. We
define
\[k_{\Delta t}(t)\eqdef\floor{\frac{t}{\Delta t}} \quad \mbox{and} \quad
\kmax\eqdef k_{\Delta t}(\tmax).\]

Now, the compact domain~$\Omega\times[0,\tmax]$ is approximated by the regular
discrete grid defined by
\begin{equation}
  \forall k \in [0..\kmax],\,
  \forall i \in [0..\imax], \quad
  \bfx_i^k \eqdef
  (x_i, t^k) \eqdef
  (\xmin + i \Delta x, k \Delta t).
\end{equation}

For a function~$q$ defined over $\Omega\times[0,\tmax]$ (resp. $\Omega$), the
notation~$q_{\rm h}$ denotes any discrete approximation of~$q$ at the points of
the grid, {\em i.e.} a discrete function over $[0..\imax]\times[0..\kmax]$
(resp. $[0..\imax]$).
By extension, the notation~$q_{\rm h}$ is also a shortcut to denote the matrix
$(q_i^k)_{0\leq i\leq \imax,0\leq k\leq\kmax}$
(resp. the vector $(q_i)_{0\leq i\leq \imax}$).
The notation~$\bar{q}_{\rm h}$ is reserved to the approximation defined on
$[0..\imax]\times[0..\kmax]$ by
\[
\bar{q}_i^k\eqdef q(\bfx_i^k) \quad \mbox{(resp. $\bar{q}_i\eqdef q(x_i)$).}
\]

\begin{figure}[ht]
  \centerline{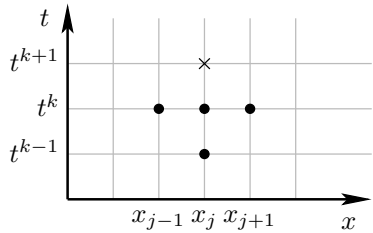}
  \caption{Three-point scheme: $p_i^{k+1}$~(at $\times$) depends on $p_{i-1}^k$,
    $p_{i}^k$, $p_{i+1}^k$, and $p_{i}^{k-1}$~(at $\bullet$).}
  \label{fig:scheme}
\end{figure}

Let~$p_{0{\rm h}}$ and~$p_{1{\rm h}}$ be two discrete functions over
$[0..\imax]$. Let~$s_{\rm h}$ be a discrete function over
$[0..\imax]\times[0..\kmax]$.
Then, the discrete function~$p_{\rm h}$ over $[0..\imax]\times[0..\kmax]$ is
said to be the solution of the three-point%
\footnote{In the sense ``three spatial points'', for the definition of
  matrix~$A_{\rm h}(c)$.}
finite difference scheme, as illustrated in Figure~\ref{fig:scheme}, when the
following set of equations holds:
\begin{multline}
  \label{e:Lh}
  \forall k \in [2..\kmax],\,
  \forall i \in [1..\imax-1], \\
  (L_{\rm h} (c) \, p_{\rm h})_i^k \eqdef
  \frac{p_i^k - 2 p_i^{k - 1} + p_i^{k - 2}}{\Delta t^2} +
  (A_{\rm h} (c) \, (i^\prime\mapsto p_{i^\prime}^{k - 1}))_i =
  s_i^{k - 1},
\end{multline}
\begin{eqnarray}
  \label{e:L1h}
  \forall i \in [1..\imax-1], & &
  (L_{1{\rm h}} (c) \, p_{\rm h})_i \eqdef
  \frac{p_i^1 - p_i^0}{\Delta t} +
  \frac{\Delta t}{2} (A_{\rm h} (c) \, (i^\prime\mapsto p_{i^\prime}^0))_i =
  p_{1,i}, \\
  \label{e:L0h}
  \forall i \in [1..\imax-1], & &
  (L_{0{\rm h}} \, p_{\rm h})_i \eqdef
  p_i^0 =
  p_{0,i}, \\
  \label{e:dirh}
  \forall k \in [0..\kmax], & &
  p_0^k =
  p_{\imax}^k =
  0
\end{eqnarray}
where the matrix~$A_{\rm h}(c)$, a discrete analog of~$A(c)$, is defined for any
vector~$q_{\rm h}$, by
\begin{equation}
  \label{e:Ah}
  \forall i \in [1..\imax-1], \quad
  \left( A_{\rm h} (c) \, q_{\rm h} \right)_i \eqdef
  -c^2 \frac{q_{i+1} - 2 q_i + q_{i-1}}{\Delta x^2}.
\end{equation}
A discrete analog of the energy is also defined by%
\footnote{By convention, the energy is defined between steps $k$
  and $k+1$, hence the notation $k+\demi$.}
\begin{equation}
  \label{e:discreteenergy}
  E_{\rm h} (c) (p_{\rm h}) ^ {k+\demi} \eqdef
  \demi \ndx{\left(i\mapsto\frac{p_i^{k+1} - p_i^k}{\Delta t}\right)}^2 +
  \demi \psAh{(i\mapsto p_i^k)}{(i\mapsto p_i^{k+1})}
\end{equation}
where, for any vectors~$q_{\rm h}$ and~$r_{\rm h}$,
\[\begin{array}{lclclcl}
\psdx{q_{\rm h}}{r_{\rm h}}&\eqdef&\sum_{i=0}^{\imax} q_ir_i\Delta x, & \quad &
\ndx{q_{\rm h}}^2&\eqdef&\psdx{q_{\rm h}}{q_{\rm h}},\\
\psAh{q_{\rm h}}{r_{\rm h}}&\eqdef&
\psdx{A_{\rm h}(c)\,q_{\rm h}}{r_{\rm h}}, & \quad &
\nAh{q_{\rm h}}^2&\eqdef&\psAh{q_{\rm h}}{q_{\rm h}}.
\end{array}\]

Note that the three-point scheme is parameterized by the discrete Cauchy
data~$p_{0{\rm h}}$ and~$p_{1{\rm h}}$, and by the discrete source
term~$s_{\rm h}$.
Of course, when these discrete inputs are respectively approximations of the
continuous functions~$p_0$, $p_1$, and~$s$ (\emph{e.g.} when
$p_{0{\rm h}}=\bar{p}_{0{\rm h}}$, $p_{1{\rm h}}=\bar{p}_{1{\rm h}}$, and
$s_{\rm h}=\bar{s}_{\rm h}$), then the discrete solution~$p_{\rm h}$ is an
approximation of the continuous solution~$p$.

\subsection{Convergence}
\label{sec:wave_conv}

Let~$\xi$ be in $]0,1[$.
The CFL($\xi$) condition (for Courant-Friedrichs-Lewy, see~\cite{cfl:pde:67})
states that the discretization steps satisfy the relation
\begin{equation}
  \label{eq:cfl}
  \frac{c \Delta t}{\Delta x} \leq 1 - \xi.
\end{equation}

The convergence error~$e_{\rm h}$ measures the distance between the continuous
and discrete solutions.
It is defined by
\begin{equation}
  \label{e:conv_error}
  \forall k \in [0..\kmax],\,
  \forall i \in [0..\imax], \quad
  e_i^k \eqdef \bar{p}_i^k - p_i^k.
\end{equation}

Note that when $p_{0{\rm h}}=\bar{p}_{0{\rm h}}$, then for all~$i$, $e_i^0=0$.
The truncation error~$\eps_{\rm h}$ measures at which precision the continuous
solution satisfies the numerical scheme.
It is defined for $k \in [2..\kmax]$ and $i \in [1..\imax-1]$ by
\begin{eqnarray}
  \label{e:trunc_error_k}
  \eps_i^k &\eqdef& (L_{\rm h} (c) \, \bar{p}_{\rm h})_i^k - \bar{s}_i^{k - 1}, \\
  \label{e:trunc_error_1}
  \eps_i^1 &\eqdef& (L_{1{\rm h}} (c) \, \bar{p}_{\rm h})_i - \bar{p}_{1,i}, \\
  \label{e:trunc_error_0}
  \eps_i^0 &\eqdef& (L_{0{\rm h}} \bar{p}_{\rm h})_i - \bar{p}_{0,i}.
\end{eqnarray}

Again, note that when $p_{0{\rm h}}=\bar{p}_{0{\rm h}}$ and
$p_{1{\rm h}}=\bar{p}_{1{\rm h}}$, then for all~$i$, $\eps_i^0=0$ and
$\eps_i^1=e_i^1/\Delta t$.
Furthermore, when there is also $s_{\rm h}=\bar{s}_{\rm h}$, then the
convergence error~$e_{\rm h}$ is itself solution of the same numerical scheme
with inputs defined by, for all~$i,k$,
\[p_{0,i}=\eps_i^0=0,\quad p_{1,i}=\eps_i^1=\frac{e_i^1}{\Delta t},\quad
\mbox{and}~s_i^k=\eps_i^{k+1}.\]

The numerical scheme is said to be convergent of order~2 if the convergence
error tends toward zero at least as fast as $\Delta x^2+\Delta t^2$ when both
discretization steps tend toward zero.\footnote{Note that~$\Delta x$
tending toward~0 actually means that~$\imax$ goes to infinity.}
More precisely, the numerical scheme is said to be convergent of
order~($m$,$n$) uniformly on the interval $[0,\tmax]$ if the convergence
error satisfies\footnote{See Section~\ref{sec:o} for the precise
  definition of the big~O notation.}
\begin{equation}
  \label{e:convergence}
  \ndx{\left(i \mapsto e_i^{k_{\Delta t}(t)}\right)} =
  O_{[0,\tmax]} (\Delta x^m + \Delta t^n).
\end{equation}

The numerical scheme is said to be consistent with the continuous problem at
order~2 if the truncation error tends toward zero at least as fast as
$\Delta x^2+\Delta t^2$ when the discretization steps tend toward~0.
More precisely, the numerical scheme is said to be consistent with the
continuous problem at order~($m$, $n$) uniformly on interval $[0,\tmax]$ if
the truncation error satisfies
\begin{equation}
  \label{e:consistency}
  \ndx{\left(i \mapsto \eps_i^{k_{\Delta t}(t)}\right)} =
  O_{[0,\tmax]} (\Delta x^m + \Delta t^n).
\end{equation}

The numerical scheme is said to be stable if the discrete solution of the
associated homogeneous problem ({\em i.e.} without any source term, $s(x,t)=0$)
is bounded independently of the discretization steps.
More precisely, the numerical scheme is said to be stable uniformly on
interval $[0,\tmax]$ if the discrete solution of the problem without any
source term satisfies
\begin{multline}
  \label{e:stability}
  \exists \alpha, C_1, C_2 > 0,\,
  \forall t \in [0, \tmax],\,
  \forall \Delta x, \Delta t > 0, \quad
  \sqrt{\Delta x^2 + \Delta t^2} < \alpha \Rightarrow \\
  \ndx{\left(i \mapsto p_i^{k_{\Delta t}(t)}\right)} \leq
  (C_1 + C_2 t) (\ndx{p_{0{\rm h}}} + \nAh{p_{0{\rm h}}} + \ndx{p_{1{\rm h}}}).
\end{multline}

The result to be formally proved at Section~\ref{sec:wave_proof} states that if
the continuous solution~$p$ is regular enough on~$\Omega\times[0,\tmax]$ and if
the discretization steps satisfy the CFL($\xi$) condition, then the three-point
scheme is convergent of order (2, 2) uniformly on interval $[0,\tmax]$.

We do not admit (nor prove) the Lax equivalence theorem which stipulates that
for a wide variety of problems and numerical schemes, consistency implies the
equivalence between stability and convergence.
Instead, we establish that consistency and stability implies convergence in
the particular case of the one-dimensional acoustic wave equation.

\subsection{Program}
\label{sec:program}

\begin{lstlisting}[
    float,
    caption={The main part of the {\clang} code, without annotations.},
    label={l:unannotated_code},
    language=C,
    numbers=left, numberstyle=\tiny, stepnumber=5, firstnumber=0,
    frame=trBL
  ]
/* Compute the constant coefficient of the stiffness matrix. */
a1 = dt/dx*v;
a  = a1*a1;

/* First initial condition and boundary conditions. */
/* Left boundary. */
p[0][0] = 0.;
/* Time iteration -1 = space loop. */
for (i=1; i<ni; i++) {
  p[i][0] = p0(i*dx);
}
/* Right boundary. */
p[ni][0] = 0.;

/* Second initial condition (with p1=0) and boundary conditions. */
/* Left boundary. */
p[0][1] = 0.;
/* Time iteration 0 = space loop. */
for (i=1; i<ni; i++) {
  dp = p[i+1][0] - 2.*p[i][0] + p[i-1][0];
  p[i][1] = p[i][0] + 0.5*a*dp;
}
/* Right boundary. */
p[ni][1] = 0.;

/* Evolution problem and boundary conditions. */
/* Propagation = time loop. */
for (k=1; k<nk; k++) {
  /* Left boundary. */
  p[0][k+1] = 0.;
  /* Time iteration k = space loop. */
  for (i=1; i<ni; i++) {
    dp = p[i+1][k] - 2.*p[i][k] + p[i-1][k];
    p[i][k+1] = 2.*p[i][k] - p[i][k-1] + a*dp;
  }
  /* Right boundary. */
  p[ni][k+1] = 0.;
}
\end{lstlisting}

The main part of the C program is listed in
Listing~\ref{l:unannotated_code}.

The grid steps $\Delta x$ and $\Delta t$ are respectively represented by the
variables {\code dx} and {\code dt}, the grid sizes $\imax$ and $\kmax$ by the
variables {\code ni} and {\code nk}, and the propagation velocity $c$ by the
variable {\code v}.
The initial position $p_{0{\rm h}}$ is represented by the function {\code p0}.
The initial velocity $p_{1{\rm h}}$ and the source term $s_{\rm h}$ are
supposed to be zero and are not represented.
The discrete solution $p_{\rm h}$ is represented by the two-dimensional array
{\code p} of size $(\imax+1) \times (\kmax+1)$.
(This is a simple naive implementation, a more efficient implementation would
store only two time steps.)

To assemble the stiffness matrix $A_{\rm h}(c)$, we only have to compute the
square of the CFL coefficient $\frac{c\Delta t}{\Delta x}$ (lines 1--2).
Then, we recognize the space loops for the initial conditions:
Equation~(\ref{e:L0h}) on lines 6--8, and Equation~(\ref{e:L1h}) with
$p_{1{\rm h}}=0$ on lines 14--17.
The space-time loop on lines 23--28 for the evolution problem comes from
Equation~(\ref{e:Lh}).
And finally, the boundary conditions of Equation~(\ref{e:dirh}) are spread out
on lines 9--10, 18--19, and 29--30.


\section[Bounding errors]{Bounding Errors}\label{sec:errors}

\subsection[Method error]{Method Error}
\label{sec:wave_proof}

We first present the notions necessary to formalize and prove the method
error.
Then, we detail how the proof is conducted:
we establish the consistency, the stability and prove that these two properties imply convergence in the case of the one-\-dimensional acoustic wave
equation.

\subsubsection[Big~O, differentiability, and regularity]{Big~O, Differentiability, and Regularity}
\label{sec:o}

When considering a big~O equality $a = O(b)$, one usually assumes that
$a$ and~$b$ are two expressions defined over the same domain and its
interpretation as a quantified formula comes naturally. Here the
situation is a bit more complicated. Consider
\[f(\bfx,\bfdeltax) = O(g(\bfdeltax))\]
when $\norm{\bfdeltax}$ goes to~0.
If one were to assume that the equality holds for any
$\bfx \in \mathbb{R}^2$, one would interpret it as
\[\forall \bfx, \exists \alpha > 0, \exists C > 0,\forall \bfdeltax, \quad
\norm{\bfdeltax} \le \alpha \Rightarrow |f(\bfx,\bfdeltax)| \le C \cdot
|g(\bfdeltax)|,\]
which means that constants $\alpha$ and $C$ are in fact functions of
$\bfx$. Such an interpretation happens to be useless, since the infimum
of $\alpha$ may well be zero while the supremum of $C$ may be $+\infty$.

A proper interpretation requires the introduction of a uniform
big~O relation with respect to the additional variable $\bfx$:
\begin{multline}
  \label{eq:Oups}
  \exists \alpha > 0,\, \exists C > 0,\,
  \forall \bfx \in \Omega_\bfx,\, \forall \bfdeltax \in \Omega_\bfdeltax,
  \mymultlinenewline
  \norm{\bfdeltax} \le \alpha \Rightarrow |f(\bfx,\bfdeltax)| \le
  C \cdot |g(\bfdeltax)|.
\end{multline}

To emphasize the dependency on the two subsets $\Omega_\bfx$ and
$\Omega_\bfdeltax$, uniform big~O equalities are now written
\[
f(\bfx,\bfdeltax) = O_{\Omega_\bfx, \Omega_\bfdeltax}(g(\bfdeltax)).
\]

We now precisely define the notion of ``sufficiently regular'' functions in
terms of the full-fledged notation for the big~O.
The further result on the convergence of the numerical scheme requires that
the solution of the continuous equation is actually sufficiently regular.
We introduce two operators that, given a real-valued function $f$ defined on
the 2D plane and a point in the plane, return the values
$\frac{\partial f}{\partial x}$ and $\frac{\partial f}{\partial t}$ at this
point. Given these two operators, we can define
the usual 2D Taylor polynomial of order~$n$ of a function~$f$:
\[\mathrm{TP}_n(f,\bfx) \eqdef (\Delta x,\Delta t) \mapsto \sum_{p=0}^n
\frac{1}{p!} \left( \sum_{m=0}^p \binom{p}{m} \cdot \frac{\partial^p
f}{\partial x^m \partial t^{p-m}}(\bfx) \cdot \Delta x^m \cdot \Delta
t^{p-m} \right).\]

Let $\Omega_\bfx\subset\R^2$.
We say that the previous Taylor polynomial is a uniform approximation of
order~$n$ of~$f$ on $\Omega_\bfx$ when the following uniform big~O equality
holds:
\[
f(\bfx + \bfdeltax) - \mathrm{TP}_n(f,\bfx)(\bfdeltax) =
O_{\Omega_\bfx, \R^2}\left(\norm{\bfdeltax}^{n+1}\right).
\]

A function~$f$ is then said to be {\em sufficiently regular of order~$n$
uniformly on $\Omega_\bfx$} when all its Taylor polynomials of order smaller
than~$n$ are uniform approximations of~$f$ on $\Omega_\bfx$.

\subsubsection{Consistency}
\label{sec:consist}

The consistency of a numerical scheme expresses that, for $\bfdeltax$
small enough, the continuous solution taken at the points of the grid almost
solves the numerical scheme.
More precisely, we formally prove that when the continuous solution of the wave
equation~(\ref{e:L})--(\ref{e:dir}) is sufficiently regular of order~4
uniformly on $[\xmin,\xmax]\times[0,\tmax]$, the numerical
scheme~(\ref{e:Lh})--(\ref{e:dirh}) is consistent with the continuous problem
at order~(2, 2) uniformly on interval $[0,\tmax]$ (see
definition~(\ref{e:consistency}) in Section~\ref{sec:wave_conv}).
This is obtained using the properties of Taylor approximations;
the proof is straightforward while involving long and complex expressions.

The key idea is to always manipulate uniform Taylor approximations that will be
valid for all points of all grids when the discretization steps goes down to
zero.

For instance, to take into account the initialization phase corresponding to
Equation~(\ref{e:L1h}), we have to derive a uniform Taylor approximation of
order~1 for the following continuous function (for any~$v$ sufficiently regular
of order~3)
\begin{multline*}
((x, t), (\Delta x, \Delta t)) \mapsto \mymultlinenewline
\frac{v (x, t + \Delta t) -
      v (x, t)}{\Delta t} -
    \frac{\Delta t}{2} c^2 \frac{v (x + \Delta x, t) -
      2 v (x, t) +
      v (x - \Delta x, t)}{\Delta x^2}.
\end{multline*}

Note that the expression of this function involves both $x+\Delta x$ and
$x-\Delta x$, meaning that we need a Taylor approximation which is valid for
both positive and negative growths.
The proof would have been impossible if we had required $0 < \Delta x$ (as a
space grid step) in the definition of the Taylor approximation.

In contrast with the case of an infinite string \cite{BCFMMW10}, we
do not need here a lower bound for $c \frac{\Delta t}{\Delta x}$.

\subsubsection{Stability}
\label{sec:stab}

The stability of a numerical scheme expresses that the growth of the discrete
solution is somehow bounded in terms of the input data (here, the Cauchy data
$u_{0{\rm h}}$ and $u_{1{\rm h}}$, and the source term $s_{\rm h}$).
For the proof of the round-off error (see Section~\ref{sec:round}), we need a
statement of the same form as definition~(\ref{e:stability}) of
Section~\ref{sec:wave_conv}.
Therefore, we formally prove that, under the CFL$(\xi)$
condition~(\ref{eq:cfl}), the numerical scheme~(\ref{e:Lh})--(\ref{e:dirh}) is
stable uniformly on interval $[0,\tmax]$.

But, as we choose to prove the convergence of the numerical scheme by using an
energetic technique%
\footnote{The popular alternative, using the Fourier transform, would have
  required huge additional Coq developments.},
it is more convenient to formulate the stability in terms of the discrete
energy.
More precisely, we also formally prove that under the CFL$(\xi)$
condition~(\ref{eq:cfl}), the discrete energy~(\ref{e:discreteenergy})
satisfies the following overestimation,
$$\sqrt{E_{\rm h} (c) (p_{\rm h})^{k + \demi}} \le
\sqrt{E_{\rm h} (c) (p_{\rm h})^{\demi}} + \frac{\sqrt{2}}{2\sqrt{2 \xi -
    \xi^2}} \cdot \Delta t \cdot \sum_{k^\prime=1}^k
    \norme{\Delta x}{\left(i \mapsto s_i^{k^\prime}\right)}$$
for all $t\in[0,\tmax]$ and with
$k=\left\lfloor\frac{t}{\Delta t}\right\rfloor-1$.

The evolution of the discrete energy between two consecutive time
steps is shown to be proportional to the source term.
In particular, the energy is constant when the source is inactive.
Then, we obtain the following underestimation of the discrete energy,
$$  \forall k, \quad
   \demi\left(1-\left(c \frac{\Delta t}{\Delta x}\right)^2\right)
  \norme{\Delta x}{\left(i \mapsto \frac{p_i^{k + 1} - p_i^k}{\Delta t}\right)}
  \le {E_{\rm h} (c) (p_{\rm h})^{k + \demi}}.$$
Therefore, the non-negativity of the discrete energy is directly related to the
CFL$(\xi)$ condition.

Note that this stability result is valid for any input data $p_{0{\rm h}}$,
$p_{1{\rm h}}$, and $s_{\rm h}$.

\subsubsection{Convergence}
\label{sec:conv}

The convergence of a numerical scheme expresses the fact that the discrete
solution gets closer to the continuous solution as the discretization steps
go down to zero.
More precisely, we formally prove that when the continuous solution of the wave
equation~(\ref{e:L})--(\ref{e:dir}) is sufficiently regular of order~4
uniformly on $[\xmin,\xmax]\times[0,\tmax]$, and under the CFL$(\xi)$
condition~(\ref{eq:cfl}), the numerical scheme~(\ref{e:Lh})--(\ref{e:dirh}) is
convergent of order~(2, 2) uniformly on interval $[0,\tmax]$ (see
definition~(\ref{e:convergence}) in Section~\ref{sec:wave_conv}).

Firstly, we prove that the convergence error $e_{\rm h}$ is itself the discrete
solution of a numerical scheme of the same form but with different input data%
\footnote{Of course, there is no associated continuous problem.}.
In particular, the source term (on the right-hand side) is here the truncation
error $\eps_{\rm h}$ associated with the initial numerical scheme for $p_{\rm h}$.
Then, the previous stability result holds, and we have an overestimation of the
square root of the discrete energy associated with the convergence error
$E_{\rm h}(c)(e_{\rm h})$ that involves a sum of the corresponding source
terms, {\em i.e.} the truncation error.
Finally, the consistency result also makes this sum a big~O of
$\Delta x^2+\Delta t^2$, and a few more technical steps conclude the proof.


\subsection[Round-off error]{Round-off Error}
\label{sec:round}

As each operation is done with IEEE-754 floating-point numbers~\cite{ieee-754},
round-off errors will occur and may endanger the accuracy of the final
results. On this program, naive forward error analysis
gives an error bound that is proportional to $2^k 2^{-53}$ for the
computation of a $p_i^k$. If this bound was sensible, it would cause the
numerical scheme to compute only noise after a few steps. Fortunately,
round-off error actually compensate themselves. To take into account the
compensations and hence prove a usable error bound, we need a precise
statement of the round-off error \cite{Bol09b} to exhibit the
cancellations made by the numerical scheme.

\newcommand{\ul}{\underline}

\subsubsection[Local round-off errors]{Local Round-off Errors}
\label{sub:delta-seq}

Let $\delta_i^k$ be the (signed) floating-point error made in the
two lines computing $p_i^k$ (lines 26--27 in
Listing~\ref{l:unannotated_code}). Floating-point values as computed by the
program will be underlined: $\ul{a}$, $\ul{p}_i^k$
to distinguish them from the discrete values of
previous sections. They match the expressions
{\code a} and {\code p[i][k]} in the annotations, while $a$ and
$p_i^k$ can be
represented in the annotations by {\code $\backslash$exact(a)} and
{\code $\backslash$exact(p[i][k])}, as described in Section \ref{tool:fp}.

The $\delta_i^k$ are defined as follow:
\[\delta_i^{k+1} = \ul{p}_i^{k+1} - (2 \ul{p}_i^k - \ul{p}_i^{k-1} +
a \times (\ul{p}_{i+1}^k - 2 \ul{p}_i^k + \ul{p}_{i-1}^k)).\]

Note that the program explained in Section~\ref{sec:program} gives us
that
\[\ul{p}_i^{k+1} = \mbox{fl}\left(2 \ul{p}_i^k - \ul{p}_i^{k-1} +
\ul{a} \times (\ul{p}_{i+1}^k - 2 \ul{p}_i^k + \ul{p}_{i-1}^k)\right)\]
where $\mbox{fl}(\cdot)$ means that all the arithmetic operations that
appear between the parentheses are actually performed by floating-point
arithmetic, hence a bit off.

In order to get a bound on $\delta_i^k$, we need to have the range of
$\ul{p}_i^k$. For this bound to be usable in our correctness proof, we
need the range to be $[-2,2]$. We have proved this fact by
using the bounds on the method error and the round-off error of all the
$\ul{p}^k$ and $\ul{p}^{k-1}$.

To prove the bound on $\delta_i^k$, we perform forward error analysis and
then use interval arithmetic to bound each intermediate error. We
prove that, for all $i$ and $k$, we have
 $|\delta_i^{k}| \le 78 \times 2^{-52}$ for a
reasonable error bound for $a$, that is to say $|\ul{a} - a| \le
2^{-49}$.

\subsubsection[Convolution of round-off errors]{Convolution of Round-off Errors}

Note that the global floating-point error $\Delta_i^k= \ul{p}_i^k -
p_i^k$ depends not only on $\delta_i^k$, but also on all the
$\delta_{i+j}^{k-l}$ for $0 < l \le k$ and $-l \le j \le l$. Indeed
round-off errors propagate along floating-point computations. Their
contributions to $\Delta_i^k$, which are independent and linear (due to
the structure of the numerical scheme), can be computed by performing a
convolution with a function $\lambda: (\Z \times \Z) \ra \R$. This
function $\lambda$ represents the results of the numerical scheme when
fed with a single unit value:
\[\begin{array}{c}
\lambda_0^0=1 \qquad \forall i \neq 0, \ \lambda_i^0=0\\[0.5em]
\lambda_{-1}^1=\lambda_1^1 = a \qquad \lambda_0^1=2 (1-a)
\qquad \forall i \not \in \{-1,0,1\},  \ \lambda_i^1=0\\[0.5em]
\lambda_i^k = a \times (\lambda_{i-1}^{k-1} + \lambda_{i+1}^{k-1}) +
2 (1-a) \times \lambda_i^{k-1} - \lambda_i^{k-2}
\end{array}\]

\begin{theorem}
\label{th:delta}
\[\Delta_i^k = \ul{p}_i^k - p_i^k = \sum_{l=0}^k \sum_{j=-l}^l
\lambda_j^l \ \delta_{i+j}^{k-l}.\]
\end{theorem}

Details of the proof can be found in \cite{Bol09b}, but this point of view
using convolution is new. The
proof mainly amounts to performing numerous tedious transformations of
summations until both sides are proved equal.

The previous proof assumes that the double summation is correct for
all $(i',k')$ such that $k' < k$. This would be correct if there was
an unbounded set of $i$ where $p_i^k$ is computed. This is no longer the
case for a finite string. Indeed, at the ends of the range ($i = 0$ or
$\imax$), $p_i^k$ and $\ul{p}_i^k$ are equal to $0$, so $\Delta_i^k$ has
to be $0$ too.

The solution is to define the successive antisymmetric extension in space
(as is done for d'Alembert's formula in Section~\ref{sec:continuous}) and
to use it instead of $\delta$. This ensures that both $\Delta_0^k$ and
$\Delta_{\imax}^k$ are equal to $0$. It does not have any consequence on
the values of $\Delta_i^k$ for $0 < i < \imax$.

\subsubsection[Bound on the global round-off error]{Bound on the Global Round-off Error}

The analytic expression of $\Delta_i^k$ can be used to obtain a bound on
the round-off error. We will need two lemmas for this purpose.

\begin{lemma} \label{l1}
$\displaystyle \sum_{i=-\infty}^{+\infty} \lambda_i^k=k+1.$
\end{lemma}

\begin{myproof}
We have
\[\sum_{i=-\infty}^{+\infty}\!\!\!\lambda_i^{k+1}
 = 2 \check{a} \sum_{i=-\infty}^{+\infty}\!\!\!\lambda_i^k
  + 2 (1-\check{a}) \sum_{i=-\infty}^{+\infty}\!\!\!\lambda_i^k -
  \sum_{i=-\infty}^{+\infty}\!\!\!\lambda_i^{k-1}
 = 2 \sum_{i=-\infty}^{+\infty}\!\!\!\lambda_i^k-
  \sum_{i=-\infty}^{+\infty}\!\!\!\lambda_i^{k-1}.\]

The sum by line verifies a simple linear recurrence.
As $\sum \lambda_i^0=1$ and $\sum \lambda_i^1=2$, we have
$\sum \lambda_i^k=k+1$.
\end{myproof}

\begin{lemma}
\label{th:jacobi}
 $\lambda_i^k \ge 0$.
\end{lemma}

\begin{myproof}
The demonstration was found out by M. Kauers and V. Pillwein.

If we denote by $P$ the Jacobi polynomial, we have
$$\lambda_n^j =
  \sum_{k=j}^n \binom{2k}{j+k} \binom{n+k+1}{2k+1} (-1)^{j+k} a^k
  = a^j \sum_{k=0}^{n-j}  P_k^{(2j,0)}(1-2a)$$

Now the conjecture follows directly from the inequality of Askey and
Gasper~\cite{AsGa72}, which asserts that $\sum_{k=0}^n
P_k^{(r,0)}(x) > 0$ for $r>-1$ and $-1<x\leq 1$ (see Theorem 7.4.2 in The
Red Book~\cite{AnAsRo99}).
\end{myproof}

\begin{theorem}
\label{th:roundoff_error}
\[\left| \Delta_i^k \right| = \left| \ul{p}_i^k - p_i^k \right|
\le 78  \times 2^{-53} \times (k+1) \times (k+2).\]
\end{theorem}

\begin{myproof}
According to Theorem~\ref{th:delta}, $\Delta_i^k$ is equal to
$\sum_{l=0}^k \sum_{j=-l}^l \lambda_j^l \ \delta_{i+j}^{k-l}$. We know
that for all $j$ and $l$, $|\delta_j^l| \le 78 \times 2^{-52}$ and that
$\sum \lambda_i^l=l+1$. Since the $\lambda_i^k$ are nonnegative, the
error is easily bounded by $78 \times 2^{-52} \times \sum_{l=0}^k (l+1)$.
\end{myproof}

\subsection[Total error]{Total Error}
\label{sec:total_error}

Let~$\cal E_{\rm h}$ be the total error.
It is the sum of the method error (or convergence error) $e_{\rm h}$ of
Sections~\ref{sec:wave_conv} and~\ref{sec:conv}, and of the round-off
error~$\Delta_h$ of Section~\ref{sec:round}.

From Theorem~\ref{th:roundoff_error}, we can estimate%
\footnote{When $\frac{\tmax}{\Delta t}\geq 2$, we have
  $\left(\frac{\tmax}{\Delta t} + 1\right)
  \left(\frac{\tmax}{\Delta t} + 2\right) \leq
  3 \frac{\tmax^2}{\Delta t^2}$.}
the following upper bound for the spatial norm of the round-off error when
$\Delta x\leq 1$ and $\Delta t\leq\tmax/2$:
for all $t\in[0,\tmax]$,
\begin{eqnarray*}
  \ndx{\left(i \mapsto \Delta_i^{k_{\Delta t}(t)}\right)}
  & = & \sqrt{\sum_{i=0}^{\imax}
    \left( \Delta_i^{k_{\Delta t}(t)} \right)^2 \Delta x} \\
  & \le & \sqrt{(\imax+1)\Delta x} \times 78  \times 2^{-53} \times
  \left( \frac{\tmax}{\Delta t} + 1 \right) \times
  \left( \frac{\tmax}{\Delta t} + 2 \right) \\
  & \leq & \sqrt{\xmax - \xmin + 1} \times 78 \times 2^{-53} \times
  3 \times \frac{\tmax^2}{\Delta t^2}.
\end{eqnarray*}

Thus, from the triangular inequality for the spatial norm, we obtain the
following estimation of the total error:
\begin{multline*}
  \forall t \in [0, \tmax], \;
  \forall \bfdeltax, \quad
  \|\bfdeltax\| \leq \min(\alpha_e, \alpha_\Delta) \Rightarrow
  \mymultlinenewline
  \ndx{\left(i \mapsto {\cal E}_i^{k_{\Delta t}(t)}\right)} \leq
  C_e (\Delta x^2 + \Delta t^2) + \frac{C_\Delta}{\Delta t^2}
\end{multline*}
where the convergence constants~$\alpha_e$ and~$C_e$ were extracted
from the Coq proof (see Section~\ref{sec:conv}) and are given in terms of
the constants for the Taylor approximation of the exact solution at degree~3
($\alpha_3$ and $C_3$), and at degree~4 ($\alpha_4$ and $C_4$) by
\begin{eqnarray*}
  \alpha_e & = & \min (1, \tmax, \alpha_3, \alpha_4), \\
  C_e & = & 2 \mu \tmax \sqrt{\xmax - \xmin}
  \left(\frac{C^\prime}{\sqrt{2}} + \mu (\tmax + 1) C^{\prime\prime} \right)
\end{eqnarray*}
with $\mu=\frac{\sqrt{2}}{\sqrt{2\xi-\xi^2}}$,
$C^\prime=\max(1,C_3+c^2 C_4+1)$, and
$C^{\prime\prime}=\max(C^\prime,2(1+c^2)C_4)$,
and where the round-off constants~$\alpha_\Delta$ and~$C_\Delta$, as
explained above,  are given by
\begin{eqnarray*}
  \alpha_\Delta & = & \min(1, \tmax/2), \\
  C_\Delta & = & 234 \times 2^{-53} \times \tmax^2 \sqrt{\xmax - \xmin + 1}.
\end{eqnarray*}

\begin{figure}[ht]
  \begin{center}
    \includegraphics[width=0.47\linewidth]{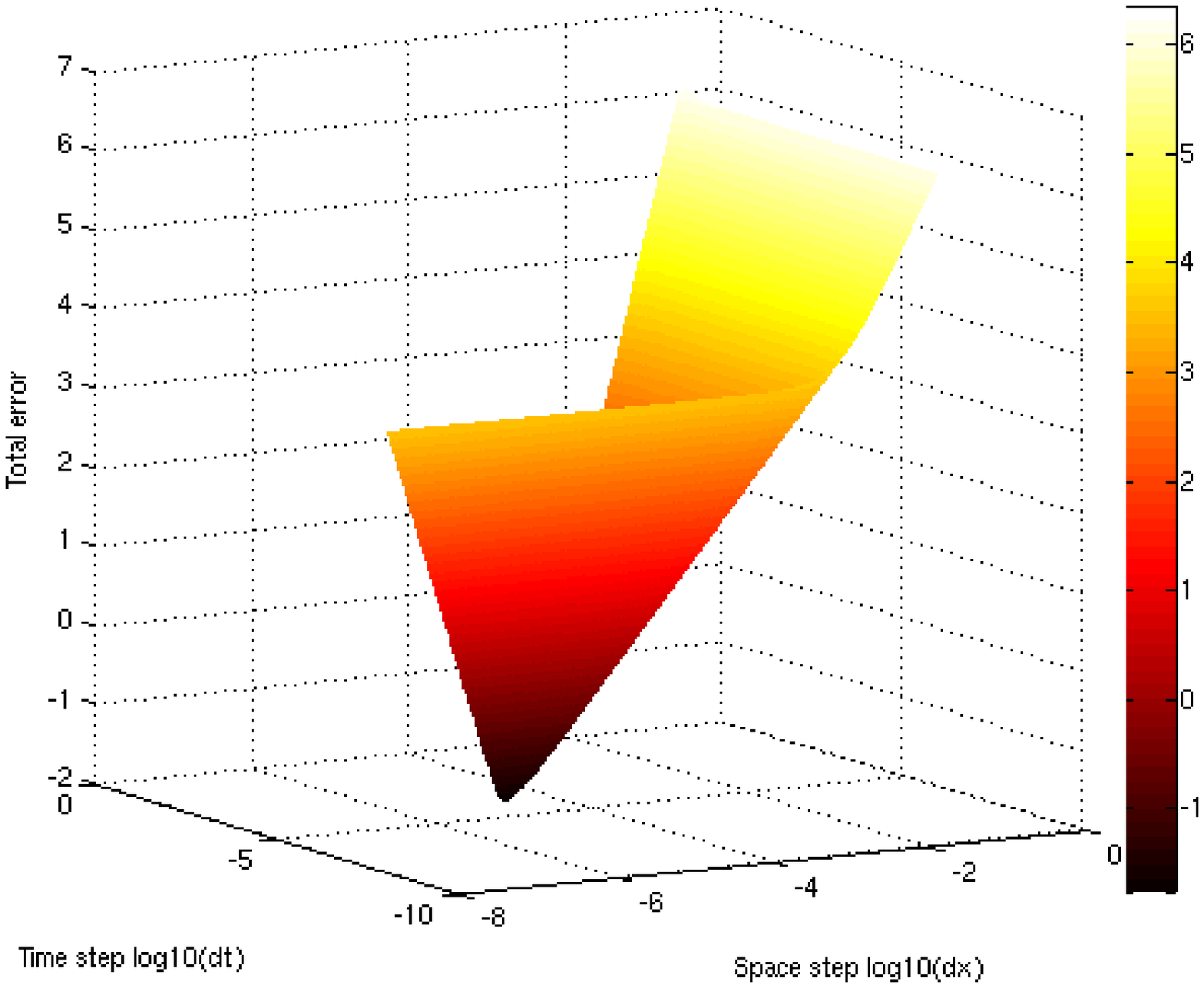}
    \hfill
    \includegraphics[width=0.47\linewidth]{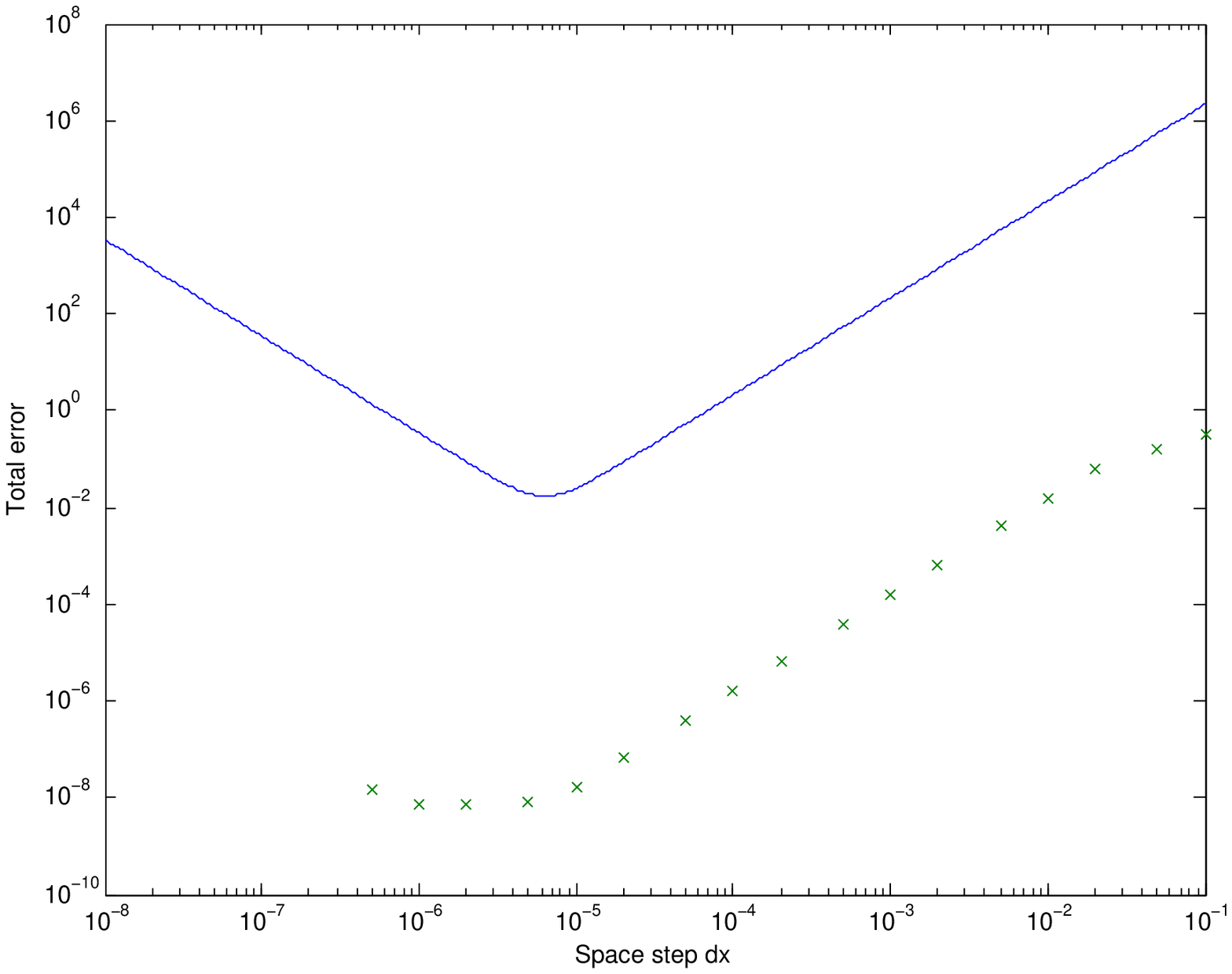}
  \end{center}
  \caption{Upper bound for the total error in log-scale.
    Left: for~$\Delta x$ and~$\Delta t$ satisfying the CFL condition.
    The lighter area (in yellow) represents the higher values above $10^4$,
    whereas the darker area represents the lower values below $10^{-1}$.
    Right: for an optimal CFL condition with
    $\Delta t=\frac{1-\xi}{c}\Delta x$.
  The green crosses represent the effective total error computed by the
  C program for a few values of the space step.}
  \label{f:total_error}
\end{figure}

To give an idea of the relative importance of both errors, we consider the
academic case where the space domain is the interval $[0,1]$, the velocity of
waves is $c=1$, and there is no initial velocity ($u_1(x)=0$) nor source term
($s(x,t)=0$).
We suppose that the initial position is given by
$u_0(x)=\chi (2(x-x_0)/l)$ where~$x_0=0.5$, $l=0.25$,
and~$\chi$ is the~$C^4$ function defined on $[-1,1]$ by
$\chi(z)=(\cos(\frac{\pi}{2}z))^5$, and with null continuation on
the real axis.
For this function, we may take $\alpha_3=\alpha_4=\sqrt{2}/2$,
$C_3=5120\sqrt{2}$, and $C_4=409600/3$.
The corresponding solution presents two hump-shaped signals that propagate in
each direction along the string, see Figure~\ref{f:waves}.

The upper bound on the total error is represented in
Figure~\ref{f:total_error}.
Note that everything is in logarithmic scale.
Of course, decreasing the size of the grid step decreases the method error,
but in the same time, it increases the round-off error.
Hence, the existence of a minimum for the upper bound on the total error (about
0.02 in our test case), corresponding to optimal grid step sizes.
Fortunately, the effective total error usually happens to be much smaller than
this upper bound (by about a factor of $10^6$ in our example).

Even if the effective total error on this example is off by several
orders of magnitude with respect to the theoretical bound, this
experiment is still reassuring. First, the left side of
Figure~\ref{f:total_error} shows that the optimal choice (the darker
part) for choosing $\Delta x$ and $\Delta t$ is reached near the limit of
the CFL condition. This property matches common knowledge from numerical
analysis. Second, the right side shows that both the effective error and
the theoretical error have the same asymptotic behavior. So the properties
we have verified in this work are not intrinsically easier than the best
theorems one could state. It is just that the constants of the formulas
extracted from the proofs (which we did not tune for this specific
purpose) are not optimal for this example.


\section[Mechanization of proofs]{Mechanization of Proofs}
\label{sec:mechanization}

In Sections~\ref{sec:wave_proof} and~\ref{sec:round}, we have mostly described
the method and round-off errors introduced when solving the wave equation
problem with the given numerical scheme.  We do not yet know whether this
formalization actually matches the program described in
Section~\ref{sec:program} and fully given in
Appendix~\ref{sec:annotated_source}. In addition, the program might contain
programming errors like out-of-bound accesses, which would possibly be left
unattended while comparing the program and its formalization.

To fully verify the program, our process is as follows. First, we
annotated the C program with comments specifying its behavioral
properties, that is, what the program is supposed to compute. Second, we
let Frama-C/Why generate proof obligations that state that the program
matches its specification and that its execution is safe. Third, we used
automated provers and Coq to prove all of these obligations.

Section~\ref{tools} presents all the tools we have used for verifying
the C program. Then Section~\ref{sec:annot} explains how the program was
annotated. Finally, Section~\ref{sec:auto} shows how we proved all the
obligations, either automatically or with a proof assistant.

\subsection{Tools}
\label{tools}

Several software packages are used in this work. The formal proof of the
method error has been made in Coq. The formal proof of the round-off
error has been made in Coq, and using the Gappa tactic. The certification
of the C program has used Frama-C (with the Jessie plug-in), and to prove
the produced goals, we used Gappa, SMT provers, and the preceding Coq
proofs. This section is devoted to present these tools and
necessary libraries.

\subsubsection{Coq}
\label{tool:coq}

Coq\footnote{\url{http://coq.inria.fr/}} is a formal system that provides
an expressive language to write mathematical definitions, executable
algorithms, and theorems, together with an interactive environment for
proving them~\cite{Coq}. Coq's formal language is based on the Calculus of Inductive
Constructions~\cite{CIC} that combines both a higher-order logic and a
richly-typed
functional programming language. Coq allows to define functions or
predicates, that can be evaluated efficiently, to state mathematical
theorems and software specifications, and to interactively develop formal
proofs of these theorems. These proofs are machine-checked by a
relatively small \emph{kernel}, and certified programs can be extracted
from them to external programming languages like Objective Caml,
Haskell, or Scheme~\cite{Let02}.

As a proof development system, Coq provides interactive proof methods,
decision and semi-decision algorithms, and a tactic language for
letting the user define its own proof methods. Connection with
external computer algebra system or theorem provers is also available.

The Coq library is structured into two parts: the initial library, which
contains elementary logical notions and data-types, and the standard library,
a general-purpose library containing various developments and
axiomatizations about sets, lists, sorting, arithmetic, real numbers,
etc.

In this work, we mainly use the Reals standard library \cite{May01}, that is
a classical axiomatization of an Archimedean ordered complete field. We chose
Reals to make our numerical proofs because we do not need an
intuitionistic formalization.

For floating-point numbers, we use a large Coq
library\footnote{\url{http://lipforge.ens-lyon.fr/www/pff/}} initially
developed in~\cite{DauRidThe01} and extended with various results afterwards \cite{Bol04b}.
It is a high-level
formalization of IEEE-754 with gradual underflow. This is expressed by
a formalization where floating-point numbers are pairs $(n,e)$
associated with real values $n\times \beta^e$. The requirements for a
number to be in the format $(e_{\min}, \beta^p)$ are
\[|n| < \beta^p \quad \mbox{and} \quad e_{\min} \le e.\]

This formalization is convenient for human interactive proofs as testified by the numerous
proofs using it. The huge number of lemmas available in the library (about 1400) makes it
suitable for a large range of applications. This library has since then
been superseded by the Flocq library~\cite{BolMel11}, but it was not yet
available at the time we proved the floating-point results of this work.

\subsubsection[Frama-C, Jessie, Why, and the SMT solvers]{Frama-C, Jessie, Why, and the SMT Solvers}

We use the Frama-C platform\footnote{\url{http://www.frama-c.cea.fr/}} to perform formal
verification of C programs at the source-code level.
Frama-C is an extensible framework that combines static
analyzers for C programs, written as plug-ins, within a single tool.
In this work, we use the Jessie plug-in for deductive verification.
C programs are annotated with behavioral contracts written using
the \emph{ANSI C Specification Language} (ACSL for short)~\cite{ACSL}.
The Jessie plug-in translates them to the Jessie
language~\cite{marche07plpv}, which is part of the Why
verification platform~\cite{filliatre07cav}.
This part of the process is responsible for translating the semantics
of C into a set of Why logical definitions (to model C types, memory heap,
etc.) and Why programs (to model C programs).
Finally, the Why platform computes verification conditions from these
programs, using
traditional techniques of weakest preconditions, and emits them to a
wide set of existing theorem provers, ranging from interactive proof assistants
to automated theorem provers. In this work, we use
the Coq proof assistant (Section~\ref{tool:coq}), SMT solvers
Alt-Ergo~\cite{conchon08entcs}, CVC3~\cite{CVC3} and Z3~\cite{Z3},
and the automated theorem prover Gappa (Section~\ref{tool:gappa}).
Details about automated and interactive proofs can be found in Section~\ref{sec:auto}.
The dataflow from C source code to theorem provers can be depicted as follows:
\begin{center}
  \includegraphics[width=0.8\textwidth]{verif_c.mps}
\end{center}

More precisely, to run the tools on a C program, we use a graphical
interface called gWhy. A screenshot is in Appendix
\ref{sec:screenshot}. In this interface, we may call one prover on one
or on many goals. We then get a graphical view of how many goals are
proved and by which prover.

In ACSL, annotations are using first-order logic.
Following the \emph{programming by contract} approach, the
specifications involve preconditions, postconditions, and loop invariants.
Contrary to other approaches focusing on run-time assertion checking,
ACSL specifications do not refer to C values and functions, even if pure, but
refer instead to purely logical symbols.
In the following contract for a function computing the square of an
integer {\code x}
\begin{lstlisting}[language=C]
  //@ ensures \result == x * x;
  int square(int x);
\end{lstlisting}
the postcondition, introduced with {\code ensures}, refers to the
return value {\code $\backslash$result} and argument {\code x}. Both are
denoting mathematical integer values, for the corresponding C values of type
{\code int}. In particular, {\code x $*$ x} cannot overflow.
Of course, one could give function {\code square} a more involved
specification that handles overflows, \emph{e.g.} with a precondition
requiring {\code x} to be small enough. Simply speaking, we can say
that C integers are reflected within specifications as mathematical
integers, in an obvious way. The translation of floating-point numbers
is more subtle and explained in Section \ref{tool:fp}.

\subsubsection{Gappa}
\label{tool:gappa}

The Gappa tool\footnote{\url{http://gappa.gforge.inria.fr/}} adapts
the interval-arithmetic paradigm to the proof of properties that occur
when verifying numerical applications~\cite{DauMel10}. The inputs are logical formulas
quantified over real numbers whose atoms are usually enclosures of
arithmetic expressions inside numeric intervals. Gappa answers whether it
succeeded in verifying it. In order to support program verification, one
can use \emph{rounding} functions inside expressions. These unary
operators take a real number and return the closest real number in a
given direction that is representable in a given binary floating-point
format. For instance, assuming that operator~$\circ$ rounds to the
nearest {\sf binary64} floating-point number, the following formula
states that the relative error of the floating-point addition is bounded:
\[\forall x,y\in\R,~\exists\eps\in\R,~|\eps| \le 2^{-53} \land
\circ(\circ(x)+\circ(y)) = (\circ(x) + \circ(y)) \times (1 + \eps).\]

Converting straight-line numerical programs to Gappa logical formulas is
easy and the user can provide additional hints if the tool were to fail
to verify a property. The tool is specially designed to handle codes that
are performing convoluted manipulations. For instance, it has been
successfully used to verify a state-of-the-art library of
correctly-rounded elementary functions~\cite{DinLauMel11}. In the current
work, Gappa has been used to check much simpler properties. (In
particular, no user hint was needed to discharge a proof
automatically.) But the length of their proofs would discourage even the
most dedicated users if they were to be manually handled. One of the
properties is the round-off error of a local evaluation of the numerical
scheme (Section~\ref{sub:delta-seq}). Other properties mainly deal with
proving that no exceptional behavior occurs while executing the program:
due to the initial values, all the computed values are sufficiently
small to never cause overflow.

The verification of some formulas requires reasonings that are so long
and intricate~\cite{DinLauMel11}, that it might cast some doubts on
whether an automatic tool actually succeeded in proving them. This is
especially true when the tool ends up proving a property stronger than
what the user expected. That is why Gappa also generates a formal
certificate that can be
mechanically checked by a proof assistant. This feature has served as the
basis for a Coq tactic for automatically solving goals related to
floating-point and real arithmetic~\cite{BolFilMel09}. The tactic reads
the current Coq goal, generates a Gappa goal, executes Gappa on it, recovers
the certificate, and converts it to a complete proof term that Coq
matches against the current goal. At this point, whether Gappa is correct
or not no longer matters: the original Coq goal is formally proved by a
complete Coq proof.

This tactic has been used whenever a
verification condition would have been directly proved by Gappa, if not
for some confusing notations or encodings of matrix elements. We just
had to apply a few
basic Coq tactics to put the goal into the proper form and then call the
Gappa tactic to discharge it automatically.

\subsubsection[Floating-point formalizations]{Floating-Point Formalizations}
\label{tool:fp}

A natural question is the link between the various representations of
floating-point numbers. We assume that the execution environment (mostly
the processor) complies with the IEEE-754 standard~\cite{ieee-754}, which
defines formats, rounding modes, and operations. The C program we
consider is compiled in an assembly code that will directly use these
formats and operations. We also assume that the compiler optimizations
preserve the visible semantics of floating-operations from the original
code, \emph{e.g.} no use of the extended registers. Such optimizations could
have been taken into account though, but at a cost~\cite{BolNgu11}.

When verifying the C program, the floating-point operations are
translated by Frama-C/\-Jessie/\-Why following some previous work by two of
the authors~\cite{BoldoFilliatre07}. A floating-point number $f$ is
modeled in the logic as a triple of real numbers $(r,e,m)$. Value $r$
simply stands for the real number that is immediately represented by
$f$; value $e$ stands for the \emph{exact} value of $f$, as obtained
if no rounding errors had occurred; finally, value $m$ stands for the
\emph{model} of $f$, which is a placeholder for the value intended to
be computed and filled by the user. The two latter values have no
existence in the program, but are useful for the specification and the verification. In
particular, they help state assertions about the rounding or the model
error of a program.  In ACSL, the three components of the model of a
floating-point number {\code f} can be referred to using {\code f},
{\code $\backslash$exact(f)}, and {\code $\backslash$model(f)},
respectively.  {\code $\backslash$round\_error(f)} is a macro for the
rounding error, that is, {\code $\backslash$abs(f -
  $\backslash$exact(f))}.

For instance, the following excerpt from our C program specifies the
error on the content of the {\code dx} variable, which represents the
grid step $\Delta x$ (see Section~\ref{sec:wave}).

\begin{lstlisting}[language=C,frame=trBL]
  dx = 1./ni;
  /*@ assert
    @   dx > 0. && dx <= 0.5 &&
    @   \abs(\exact(dx) - dx) / dx <= 0x1.p-53;
    @ */
\end{lstlisting}

Note that {\code 0x1.p-53} is a valid ACSL (and C too) literal meaning
$2^{-53}$.

Proof obligations are extracted from the annotated C program by computing
weakest preconditions and then translated to automated and interactive
provers. For SMT provers, the three fields $r$, $e$, and $m$, of
floating-point numbers are expressed as real numbers and operations on
floating-point numbers are uninterpreted relations axiomatized with basic
properties such as bounds on the rounding error or monotonicity. For
Gappa too, the fields are seen as real numbers. The tool, however, knows
about floating-point arithmetic and its relation to real arithmetic. So
floating-point operations are translated to the corresponding symbols
from Gappa.

For Coq, we use the formalization described in Section \ref{tool:coq}
with a limited precision and gradual underflow (so that subnormal
numbers are correctly translated). It is based on the real numbers of
the standard library, which are also used for the translation of the
exact and the model parts of the floating-point number.

While the IEEE-754 standard defines infinities and Not-a-Number as
floating-point values, our translation does not take them into account. This
does not compromise the correctness of the translation though, as each
operation has a precondition that raises a proof obligation to guarantee that
no exceptional events occur, such as overflow or division by zero, and
therefore no infinities nor Not-a-Number are produced by the program.

To summarize, there is one assumption about the actual arithmetic being
executed (IEEE-754 compliant and no overly aggressive optimizations from
the compiler) and three formalizations of floating-point arithmetic used
to verify the program: one used by Jessie/Why and then sent to the SMT
solvers, one used by Gappa, and one used by Coq. The combination of these
three different formalizations does not introduce any inconsistency.
Indeed, we have formally proved in Coq that Gappa's and Coq's
formalizations are equivalent for floating-point formats with limited
precision and gradual underflow, that is, IEEE-754 formats. We have also
formally proved that the Jessie/Why specifications and the properties for
SMT provers are compatible with these formalizations, including the
absence of special values (infinity or Not-a-Number) and the possibility to
disregard the upper bound on reals representing floating-point numbers.

In fact, there is a fourth formalization of floating-point arithmetic
involved, which is the one used internally by the interval computations
of Gappa for proving results about real-valued expressions. It is not
equivalent to the previous ones, since it is a multi-precision
arithmetic, but it has no influence whatsoever on the formalization that
Gappa uses for modeling floating-point properties.


\subsection[Program annotations]{Program Annotations}
\label{sec:annot}

The full annotations are given in Appendix~\ref{sec:annotated_source}.
We give here hints about how to specify this program.

There are two axiomatics. The first one corresponds to the mathematics: the
exact solution of the wave equation and its properties. It defines the
needed values (the exact solution $p$, and its initialization
$p_0$). We here assume that $s$ and $p_1$ are zero functions. It also
defines the derivatives of $p$ ($psol_1$, first derivative for the
first variable of $p$, and $psol_{11}$, second derivative for the
first variable, and $psol_2$ and $psol_{22}$ for the second variable)
as functions such that their value is the limit of $\frac{p(x+\Delta
  x)-p(x)}{\Delta x}$ when $\Delta x \rightarrow 0$. As the ACSL
annotations are only first order, these definitions are quite
cumbersome: each derivative needs 5 lines to be defined.

We also put as axioms the fact that the solution has the expected properties
(\ref{e:L}--\ref{e:dir}). The last property needed on the exact
solution is its regularity. We require it to be near its Taylor
approximations of degrees 3 and 4 on the whole interval $[\xmin,\xmax]$. For
instance, the following annotation states the property for degree~3.

\begin{lstlisting}[language=C,frame=trBL]
/*@ axiom psol_suff_regular_3:
  @   0 < alpha_3 && 0 < C_3 &&
  @   \forall real x; \forall real t; \forall real dx; \forall real dt;
  @   0 <= x <= 1 ==> \sqrt(dx * dx + dt * dt) <= alpha_3 ==>
  @   \abs(psol(x + dx, t + dt) - psol_Taylor_3(x, t, dx, dt)) <=
  @     C_3 * \abs(\pow(\sqrt(dx * dx + dt * dt), 3));
  @*/
\end{lstlisting}

The second axiomatic corresponds to the properties and loop invariant
needed by the program. For example, we require the matrix to be
separated: it means that a line of the matrix should not mix with
another line (or a modification could alter another point of the
matrix). We also state the existence of the loop invariant
{\code analytic\_error} that is needed for applying the results of
Section~\ref{sec:round}.

The initializations functions are specified, but not stated. This
corresponds firstly to the function {\code array2d\_alloc} that
initializes the matrix and {\code p\_zero} that produces an
approximation of the $p_0$ function. Our program verification is
modular: our proofs are generic with respect to $p_0$ and its
implementation.

The preconditions of the main functions are the following ones:
\begin{itemize}
\item $\imax$ and  $\kmax$ must be greater than one, but small enough
  so that $\imax+1$ and  $\kmax+1$ do not overflow;
\item the grid sizes $\bfdeltax$ must fulfill some mathematical
  conditions that are required for the convergence of the scheme;
\item the floating-point values computed for the grid sizes must be
  near their mathematical values;
\item to prevent exceptional behavior in the computation of $a$, the
  time discretization step must be greater than $2^{-1000}$ and
 $\frac{c \Delta t}{\Delta x}$  must be greater than $2^{-500}$. 
\end{itemize}

There are two postconditions, corresponding to the method and
round-off errors. See Sections~\ref{sec:wave_proof} and~\ref{sec:round} for
more details.

\subsection[Automation and manual proofs]{Automation and Manual Proofs}
\label{sec:auto}

This section is devoted to formal specifications and proofs
corresponding to the bounds proved in Section~\ref{sec:errors}. We
give some key points of the automated proofs.

\paragraph{Big O.}

In section~\ref{sec:o}, we present two interpretations of the big~O notation.
Usual mathematical pen-and-paper proofs switch from one interpretation to
the other depending on which one is the most adapted, without noticing
that they may not be equivalent. The formal development was helpful in
bringing into light the erroneous reasoning hidden by the usage of big~O
notations.
We introduced the notion of uniform big~O in~\cite{BCFMMW10} in the context of
an infinite string.
In the present paper, we consider the case of the finite string, hence for
compactness reasons, both notions are in fact equivalent.
However, we still use the more general uniform big~O notion to share most of
the proof developments between the finite and the infinite cases.
Regarding automation, a decision procedure has been
developed in~\cite{AD07}; unfortunately, those results were not
applicable since we needed a more powerful big~O.

\paragraph{Differential operators.}

As long as we were studying only the method error, we did not have to
define the differential operators nor assume anything about
them~\cite{BCFMMW10}. Their only properties appeared through their usage:
function $p$ is a solution of the partial differential equation and it is
sufficiently regular. This is no longer possible for the annotated C
program. Indeed, due to the underlying logic, the annotations have to
define $p$ as a solution of the PDE by using
first-order formulas stating differentiability, instead of second-order
formulas involving differential operators. Since the formalization of
Taylor approximations has been left unchanged, the natural way
to relate the C annotations with the Coq development is therefore to define the
operators as actual differential operators.
Note that this has forced us to introduce a small axiom. Indeed, our
definition of Taylor approximation depends on differential operators that are
total functions, while Coq's standard library defines only partial
operators. So we have assumed the existence of some total operators that
are equal to the partial ones whenever applied to differentiable
functions. The axiom states absolutely nothing about the result of these
operators for nondifferentiable functions, so no inconsistencies are
introduced this way. This is just a specific instance of Hilbert
$\varepsilon$ operator~\cite{Hilbert}, which does not make the logic
inconsistent~\cite{LW11}.

\paragraph{Method error.}

The Coq proof of the method error is about 5000-line long.
About half of it is dedicated to the wave equation and the other half is
re-usable (definition and properties of the dot product, the big O, Taylor
expansions\dots).
We formally proved without any axiom that the numerical scheme is convergent of
order 2, which is the known mathematical result.
An interesting aspect of the formal proof in Coq is that we were able
to extract the constants $\alpha$ and $C$ appearing in the big~O for the
convergence result in order to obtain their precise values.
The recursive extraction was fully automatic after
making explicit some inlining.
The mathematical expressions are given in Section~\ref{sec:total_error}.

\paragraph{Round-off errors.}

Except for Lemma~\ref{th:jacobi}, all
the proofs described in section~\ref{sec:round} have been done and
machined-checked using Coq. In particular, the proof of the bound on
$\delta_i^k$ was done automatically by calling Gappa from Coq.
Lemma~\ref{th:jacobi} is a technical detail compared
to the rest of our work, that is not worth the immense Coq
development it would require: keen results on integrals but also
definitions and results about the Legendre, Laguerre, Chebychev, and
Jacobi polynomials.

\paragraph{The program proof.}

Given the program code, the Why tool
generates 149 verification
conditions that have to be proved. While possible, proving all of them in
Coq would be rather tedious. Moreover, it would lead to a rather fragile
construct: any later modification to the code, however small it is, would
cause different proof obligations to be generated, which would then
require additional human interaction to adapt the Coq proofs. We prefer
to have automated provers (SMT solvers and Gappa) discharge as many of
them as possible, so that only the most intricate ones are left to be
proven in Coq. The following table shows how many goals are discharged
automatically and how many are left to the user.\footnote{Note that
verification conditions might be discharged by one or several automated
provers.}

\begin{center}
\begin{tabular}{|l|c|c|c|}
\hline
Prover  & Proved Behavior VC & Proved Safety VC & Total\\\hline
Alt-Ergo& 18                 &    80            & 98 \\\hline
CVC3    & 18                 &    89            & 107 \\\hline
Gappa   &  2                 &    20            & 22 \\\hline
Z3      & 21                 &    63            & 84     \\\hline\hline
\textbf{Automatically proved}
        & \textbf{23}        & \textbf{94}      & \textbf{117} \\ \hline
\textbf{Coq}& \textbf{21}    & \textbf{11}       & \textbf{32} \\ \hline\hline
Total   &  44                &  105             & 149 \\ \hline
\end{tabular}
\end{center}

On safety goals (matrix access, loop variant decrease, overflow),
automatic provers are helpful: they prove about 90 \%
of the goals. On behavior goals (loop invariant, assertion,
postcondition), automatic provers succeed for half of the
goals.
As our loop invariant involves an uninterpreted predicate, the
automatic provers cannot prove all the behavior goals (they would have
been too complicated anyway). That is why we resort to an interactive
higher-order theorem prover, namely Coq.

Coq proofs are split into two sets: first, the
mathematical
proof of convergence and second, the proofs of bounded round-off
errors and absence of runtime errors.
Appendix~\ref{sec:dependencies} displays the layout of the Coq
formalization.

The following tabular gives the compilation times of the Coq files on
a 3-GHz dual core machine.

\begin{center}
\begin{tabular}{|l|c|c|c|}
\hline
Type of proofs & Nb spec lines & Nb lines & Compilation time \\ \hline
Convergence   & 991 & 5\,275 & 42 s \\ \hline
Round-off + runtime errors & 7\,737 & 13\,175 & 32 min \\ \hline
\end{tabular}
\end{center}

Note that most theorem statements regarding round-off and runtime errors
are automatically generated (7\,321 lines out of 7\,737) by the
Frama-C/Jessie/Why framework.

The compilation time may seem prohibitive; it is mainly due to the
size of the theorems and to calls
to the {\code omega} decision procedure for Presburger arithmetic. The
difficulty does not lie in the arithmetic statement itself, but rather
in a large number of useless hypotheses. In order to reduce the
compilation time, we could manually massage the hypotheses to speed up
the procedure, but this would defeat the point of using an automatic
tactic.

\section{Conclusion}
\label{sec:conclusion}

In the end, having \emph{formally} verified the C program means that all
of the proof obligations generated by Frama-C/Jessie/Why have been
proved, either by automated tools or by Coq formal proofs. These formal
proofs depend on some axioms specific to this work: the fact about Jacobi
polynomials, the existence and regularity of a solution to the EDP, and the existence of
differential operators. The last two have been tackled by subsequent
works, which means that the only remaining Coq axiom is the one about
Jacobi polynomials.

We succeeded in verifying a C program that implements a numerical scheme
for the resolution of the one-dimensional acoustic wave equation. This
is comprised of three sets of proofs. First we formalized the wave
equation and proved the convergence of a scheme for its numerical
resolution. Second we proved that the C program behaves safely: no
out-of-bound array accesses and no overflow during floating-point
computations. Third we proved that the round-off errors are not causing
the numerical results to go astray. This is the first verification of
this kind of program that covers all its aspects, both mathematics and
implementation.

This work shows a tight synergy between researchers from applied mathematics
and logic. Three domains are intertwined here: applied mathematics for an
initial proof that was enriched and detailed upon request, computer
arithmetic for smart bounds on round-off errors, and formal methods for
machine-checking them. This may be the reason why such proofs never appeared
before, as this kind of collaboration is uncommon.

\medskip

Each proof came with its own hurdles. For ensuring the correct
behavior of the program, the most tedious point was to prove that setting
a result value did not cause other values to change, that is, that all
the lines of the matrix are properly separated. In particular, verifying
the loop invariant requires checking that, except for the new value, the
properties of the memory are preserved. An unexpectedly tedious part was
to check that the program actually complies with our mathematical model
for the numerical scheme.

Another difficulty lies in the mathematical proof itself. We based our work
on proofs found in books, courses, and articles. It appears that
pen-and-paper proofs are sometimes sketchy: they may be fuzzy about the
needed hypotheses, especially when switching quantifiers. We have also
learned that filling the gaps may cause us to go back to the drawing board
and to change the basic blocks of our formalization to make them more generic
(\emph{e.g.} devising a big~O that needs to be uniform and also generic with
respect to a property $P$).

\medskip

An unexpected side effect of having performed this formal verification in
Coq is our ability to automatically extract the constants hidden inside the
proofs.
That way, we are able to explicitly bound the total error rather than
just having the usual $O(\Delta x^2 + \Delta t^2)$ bound. In particular,
we can compare the magnitudes of the method error and round-off error and
then decide how to scale the discretization grid.

Coq could have offered us more: it would have been possible to
describe and prove the algorithm directly in Coq. The same formalism
would have been used all the way long, but we were more interested in
proving a real-life program in a real-life language. This has shown us
the difficulties lying in the memory handling for matrices. In the
end, we have a C code with readable annotations instead of a Coq
theorem and that seems more convincing to applied mathematicians.

\bigskip

For this exploratory work, we considered the simple three-point
scheme for the one-dimen\-sional wave equation. Further works involve
scaling to higher-dimension. The one-dimensional case showed us that summations
and finite support functions play a much more important role in the development
than we first expected. We are therefore moving to the SSReflect interface and
libraries for Coq~\cite{BGOP08}, so as to simplify the manipulations of these
objects in the higher-dimensional case.

This example also exhibits a major cancellation of rounding errors
and it would be interesting to see under which conditions numerical
schemes behave so well.

Another perspective is to generalize our approach to other higher-order
numerical schemes for the same equation, and to other PDEs. However, the proofs
of Section~\ref{sec:wave_proof} are entangled with particulars of the presented
problem, and would therefore have to be redone for other problems. So a more
fruitful approach would be to prove once and for all the Lax equivalence
theorem that states that consistency implies the equivalence between
convergence and stability. This would considerably reduce the amount of work
needed for tackling other schemes and equations.

\bibliography{biblio}
\bibliographystyle{plain}

\appendix
\normalsize

\clearpage
\section[Source code]{Source Code}
\label{sec:annotated_source}

\begin{lstlisting}[language=C,
        numbers=left, numberstyle=\tiny, stepnumber=5, firstnumber=0,
        frame=trBL]

/*@ axiomatic dirichlet_maths {
  @
  @ logic real c;
  @ logic real p0(real x);
  @ logic real psol(real x, real t);

  @ axiom c_pos: 0 < c;

  @ logic real psol_1(real x, real t);
  @ axiom psol_1_def:
  @   \forall real x; \forall real t;
  @   \forall real eps; \exists real C; 0 < C && \forall real dx;
  @   \abs(dx) < C ==>
  @   \abs((psol(x + dx, t) - psol(x, t)) / dx - psol_1(x, t)) < eps;

  @ logic real psol_11(real x, real t);
  @ axiom psol_11_def:
  @   \forall real x; \forall real t;
  @   \forall real eps; \exists real C; 0 < C && \forall real dx;
  @   \abs(dx) < C ==>
  @   \abs((psol_1(x + dx, t) - psol_1(x, t)) / dx - psol_11(x, t)) < eps;

  @ logic real psol_2(real x, real t);
  @ axiom psol_2_def:
  @   \forall real x; \forall real t;
  @   \forall real eps; \exists real C; 0 < C && \forall real dt;
  @   \abs(dt) < C ==>
  @   \abs((psol(x, t + dt) - psol(x, t)) / dt - psol_2(x, t)) < eps;

  @ logic real psol_22(real x, real t);
  @ axiom psol_22_def:
  @   \forall real x; \forall real t;
  @   \forall real eps; \exists real C; 0 < C && \forall real dt;
  @   \abs(dt) < C ==>
  @   \abs((psol_2(x, t + dt) - psol_2(x, t)) / dt - psol_22(x, t)) < eps;

  @ axiom wave_eq_0: \forall real x; 0 <= x <= 1 ==> psol(x, 0) == p0(x);
  @ axiom wave_eq_1: \forall real x; 0 <= x <= 1 ==> psol_2(x, 0) == 0;
  @ axiom wave_eq_2:
  @   \forall real x; \forall real t;
  @   0 <= x <= 1 ==> psol_22(x, t) - c * c * psol_11(x, t) == 0;
  @ axiom wave_eq_dirichlet_1: \forall real t; psol(0, t) == 0;
  @ axiom wave_eq_dirichlet_2: \forall real t; psol(1, t) == 0;

  @ logic real psol_Taylor_3(real x, real t, real dx, real dt);
  @ logic real psol_Taylor_4(real x, real t, real dx, real dt);

  @ logic real alpha_3; logic real C_3;
  @ logic real alpha_4; logic real C_4;

  @ axiom psol_suff_regular_3:
  @   0 < alpha_3 && 0 < C_3 &&
  @   \forall real x; \forall real t; \forall real dx; \forall real dt;
  @   0 <= x <= 1 ==> \sqrt(dx * dx + dt * dt) <= alpha_3 ==>
  @   \abs(psol(x + dx, t + dt) - psol_Taylor_3(x, t, dx, dt)) <=
  @     C_3 * \abs(\pow(\sqrt(dx * dx + dt * dt), 3));

  @ axiom psol_suff_regular_4:
  @   0 < alpha_4 && 0 < C_4 &&
  @   \forall real x; \forall real t; \forall real dx; \forall real dt;
  @   0 <= x <= 1 ==> \sqrt(dx * dx + dt * dt) <= alpha_4 ==>
  @   \abs(psol(x + dx, t + dt) - psol_Taylor_4(x, t, dx, dt)) <=
  @     C_4 * \abs(\pow(\sqrt(dx * dx + dt * dt), 4));

  @ axiom psol_le:
  @   \forall real x; \forall real t;
  @   0 <= x <= 1 ==> 0 <= t ==> \abs(psol(x, t)) <= 1;

  @ logic real T_max;
  @ axiom T_max_pos: 0 < T_max;

  @ logic real C_conv; logic real alpha_conv;
  @ lemma alpha_conv_pos: 0 < alpha_conv;
  @
  @ } */


/*@ axiomatic dirichlet_prog {
  @
  @  predicate analytic_error{L}
  @    (double **p, integer ni, integer i, integer k, double a, double dt)
  @    reads p[..][..];
  @
  @  lemma analytic_error_le{L}:
  @    \forall double **p; \forall integer ni; \forall integer i;
  @    \forall integer nk; \forall integer k;
  @    \forall double a; \forall double dt;
  @    0 < ni ==> 0 <= i <= ni ==> 0 <= k ==>
  @    0 < \exact(dt) ==>
  @    analytic_error(p, ni, i, k, a, dt) ==>
  @    \sqrt(1. / (ni * ni) + \exact(dt) * \exact(dt)) < alpha_conv ==>
  @    k <= nk ==> nk <= 7598581 ==> nk * \exact(dt) <= T_max ==>
  @    \exact(dt) * ni * c <= 1 - 0x1.p-50 ==>
  @    \forall integer i1; \forall integer k1;
  @    0 <= i1 <= ni ==> 0 <= k1 < k ==>
  @    \abs(p[i1][k1]) <= 2;
  @
  @  predicate separated_matrix{L}(double **p, integer leni) =
  @    \forall integer i; \forall integer j;
  @    0 <= i < leni ==> 0 <= j < leni ==> i != j ==>
  @    \base_addr(p[i]) != \base_addr(p[j]);
  @
  @ logic real sqr_norm_dx_conv_err{L}
  @   (double **p, real dx, real dt, integer ni, integer i, integer k)
  @   reads p[..][..];
  @ logic real sqr(real x) = x * x;
  @ lemma sqr_norm_dx_conv_err_0{L}:
  @   \forall double **p; \forall real dx; \forall real dt;
  @   \forall integer ni; \forall integer k;
  @   sqr_norm_dx_conv_err(p, dx, dt, ni, 0, k) == 0;
  @ lemma sqr_norm_dx_conv_err_succ{L}:
  @   \forall double **p; \forall real dx; \forall real dt;
  @   \forall integer ni; \forall integer i; \forall integer k;
  @   0 <= i ==>
  @   sqr_norm_dx_conv_err(p, dx, dt, ni, i + 1, k) ==
  @     sqr_norm_dx_conv_err(p, dx, dt, ni, i, k) +
  @     dx * sqr(psol(1. * i / ni, k * dt) - \exact(p[i][k]));
  @ logic real norm_dx_conv_err{L}
  @   (double **p, real dt, integer ni, integer k) =
  @   \sqrt(sqr_norm_dx_conv_err(p, 1. / ni, dt, ni, ni, k));
  @
  @ } */


/*@ requires leni >= 1 && lenj >= 1;
  @ ensures
  @   \valid_range(\result, 0, leni - 1) &&
  @   (\forall integer i; 0 <= i < leni ==>
  @    \valid_range(\result[i], 0, lenj - 1)) &&
  @   separated_matrix(\result, leni);
  @ */
double **array2d_alloc(int leni, int lenj);


/*@ requires (l != 0);
  @ ensures
  @   \round_error(\result) <= 14 * 0x1.p-52 &&
  @   \exact(\result) == p0(\exact(x));
  @ */
double p_zero(double xs, double l, double x);


/*@ requires
  @   ni >= 2 && nk >= 2 && l != 0 &&
  @   dt > 0. && \exact(dt) > 0. &&
  @   \exact(v) == c && \exact(v) == v &&
  @   0x1.p-1000 <= \exact(dt) &&
  @   ni <= 2147483646 && nk <= 7598581 &&
  @   nk * \exact(dt) <= T_max &&
  @   \abs(\exact(dt) - dt) / dt <= 0x1.p-51 &&
  @   0x1.p-500 <= \exact(dt) * ni * c <= 1 - 0x1.p-50 &&
  @   \sqrt(1. / (ni * ni) + \exact(dt) * \exact(dt)) < alpha_conv;
  @
  @ ensures
  @   \forall integer i; \forall integer k;
  @   0 <= i <= ni ==> 0 <= k <= nk ==>
  @   \round_error(\result[i][k]) <= 78. / 2 * 0x1.p-52 * (k + 1) * (k + 2);
  @
  @ ensures
  @   \forall integer k; 0 <= k <= nk ==>
  @   norm_dx_conv_err(\result, \exact(dt), ni, k) <=
  @     C_conv * (1. / (ni * ni) + \exact(dt) * \exact(dt));
  @ */
double **forward_prop(int ni, int nk, double dt, double v,
                      double xs, double l) {

  /* Output variable. */
  double **p;

  /* Local variables. */
  int i, k;
  double a1, a, dp, dx;

  dx = 1./ni;
  /*@ assert
    @   dx > 0. && dx <= 0.5 &&
    @   \abs(\exact(dx) - dx) / dx <= 0x1.p-53;
    @ */

  /* Compute the constant coefficient of the stiffness matrix. */
  a1 = dt/dx*v;
  a = a1*a1;
  /*@ assert
    @   0 <= a <= 1 &&
    @   0 < \exact(a) <= 1 &&
    @   \round_error(a) <= 0x1.p-49;
    @ */

  /* Allocate space-time variable for the discrete solution. */
  p = array2d_alloc(ni+1, nk+1);

  /* First initial condition and boundary conditions. */
  /* Left boundary. */
  p[0][0] = 0.;
  /* Time iteration -1 = space loop. */
  /*@ loop invariant
    @   1 <= i <= ni &&
    @   analytic_error(p, ni, i - 1, 0, a, dt);
    @ loop variant ni - i; */
  for (i=1; i<ni; i++) {
    p[i][0] = p_zero(xs, l, i*dx);
  }
  /* Right boundary. */
  p[ni][0] = 0.;
  /*@ assert analytic_error(p, ni, ni, 0, a, dt); */

  /* Second initial condition (with p_one=0) and boundary conditions. */
  /* Left boundary. */
  p[0][1] = 0.;
  /* Time iteration 0 = space loop. */
  /*@ loop invariant
    @   1 <= i <= ni &&
    @   analytic_error(p, ni, i - 1, 1, a, dt);
    @ loop variant ni - i; */
  for (i=1; i<ni; i++) {
    /*@ assert \abs(p[i-1][0]) <= 2; */
    /*@ assert \abs(p[i][0]) <= 2; */
    /*@ assert \abs(p[i+1][0]) <= 2; */
    dp = p[i+1][0] - 2.*p[i][0] + p[i-1][0];
    p[i][1] = p[i][0] + 0.5*a*dp;
  }
  /* Right boundary. */
  p[ni][1] = 0.;
  /*@ assert analytic_error(p, ni, ni, 1, a, dt); */

  /* Evolution problem and boundary conditions. */
  /* Propagation = time loop. */
  /*@ loop invariant
    @   1 <= k <= nk &&
    @   analytic_error(p, ni, ni, k, a, dt);
    @ loop variant nk - k; */
  for (k=1; k<nk; k++) {
    /* Left boundary. */
    p[0][k+1] = 0.;
    /* Time iteration k = space loop. */
    /*@ loop invariant
      @   1 <= i <= ni &&
      @   analytic_error(p, ni, i - 1, k + 1, a, dt);
      @ loop variant ni - i; */
    for (i=1; i<ni; i++) {
      /*@ assert \abs(p[i-1][k]) <= 2; */
      /*@ assert \abs(p[i][k]) <= 2; */
      /*@ assert \abs(p[i+1][k]) <= 2; */
      /*@ assert \abs(p[i][k-1]) <= 2; */
      dp = p[i+1][k] - 2.*p[i][k] + p[i-1][k];
      p[i][k+1] = 2.*p[i][k] - p[i][k-1] + a*dp;
    }
    /* Right boundary. */
    p[ni][k+1] = 0.;
    /*@ assert analytic_error(p, ni, ni, k + 1, a, dt); */
  }

  return p;

}
\end{lstlisting}

\clearpage
\section{Screenshot}
\label{sec:screenshot}

This is a screenshot of gWhy: we have the list of all the verification
conditions and if they are proved by the various automatic tools.
\begin{center}
\includegraphics[width=\linewidth]{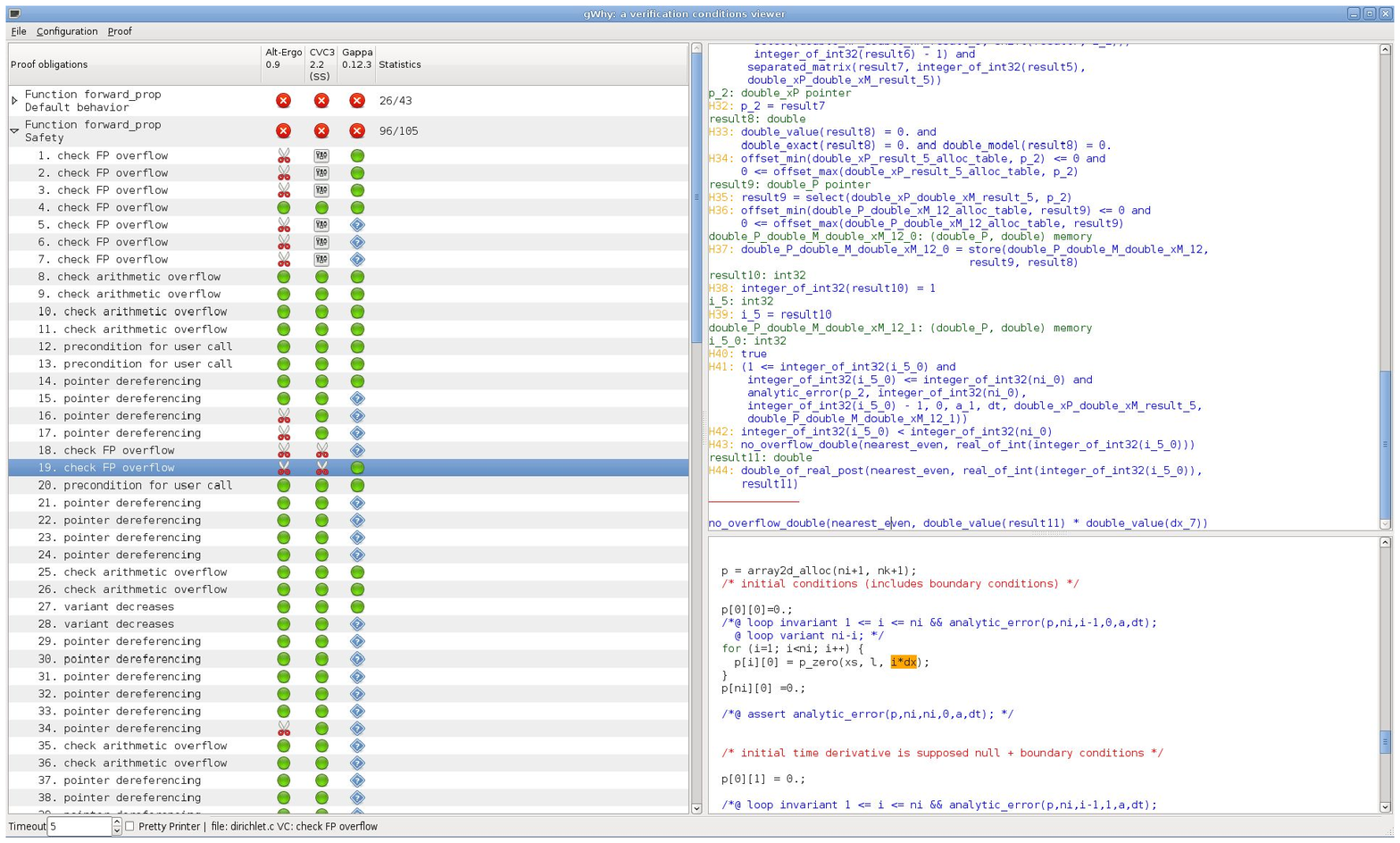}
\end{center}

\clearpage
\section[Dependency graph]{Dependency Graph}
\label{sec:dependencies}

In the following graph, the ellipse nodes are Coq files formalizing the
wave equation and the convergence of its numerical scheme. The octagon nodes
are Coq files that deal with proof obligations generated from the {\tt
dirichlet.c} program file, that is, propagation of round-off errors and
error-free execution. Arrows represent dependencies between the Coq
files.

\begin{center}
\includegraphics[width=0.6\linewidth]{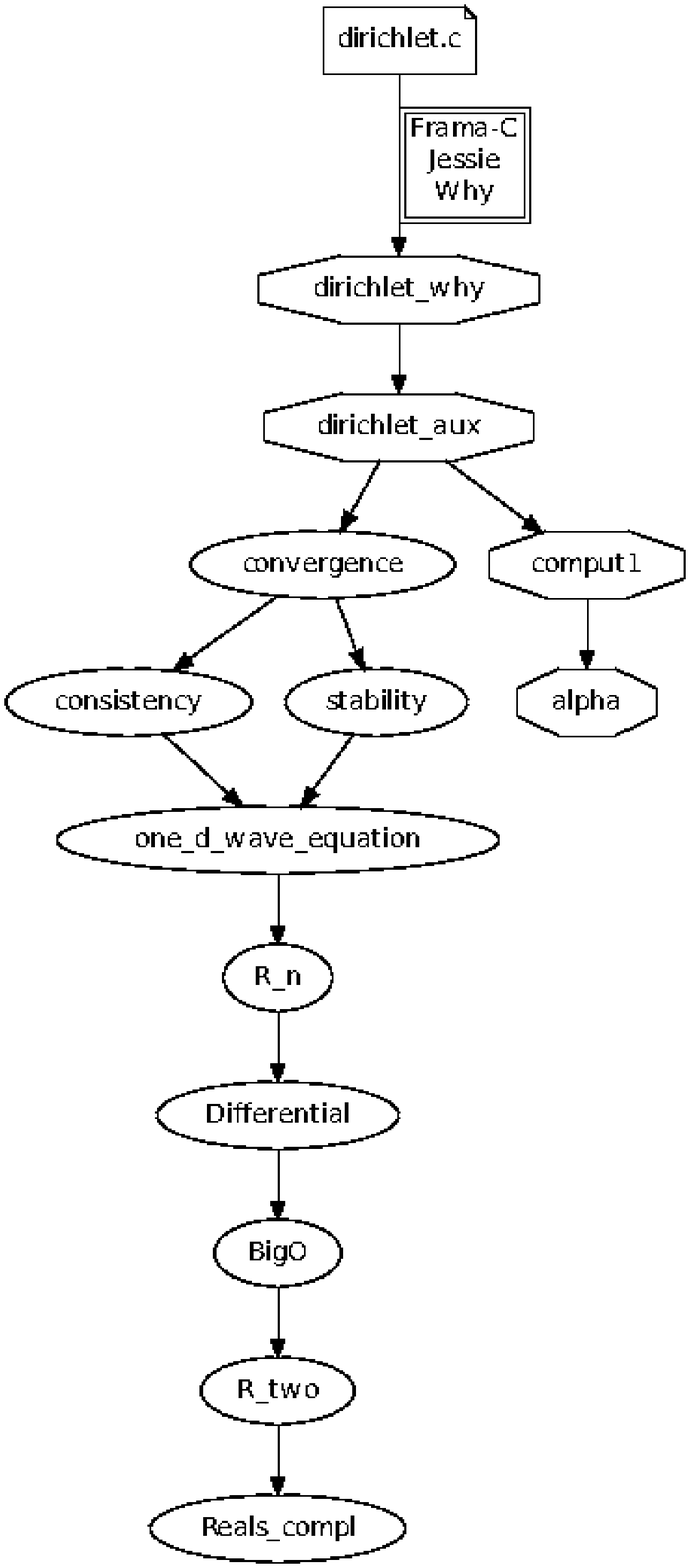}
\end{center}

\end{document}